\documentclass[a4paper,11pt]{article}
\pdfoutput=1 

\usepackage{jheppub} 

\usepackage[T1]{fontenc} 

\newcommand{\be}{\begin{equation}}
\newcommand{\ee}{\end{equation}}
\newcommand{\bea}{\begin{eqnarray}}
\newcommand{\eea}{\end{eqnarray}}
\newcommand{\ba}{\begin{array}}
\newcommand{\ea}{\end{array}}

\newcommand{\eps}{\epsilon}
\newcommand{\nn}{\nonumber}

\newcommand{\cL}{\mathcal L}
\newcommand{\cG}{\mathcal G}

\newcommand{\cA}{\mathcal A}
\newcommand{\cO}{\mathcal O}

\newcommand{\vareps}{\tilde \epsilon}
 
\def\ci{c_1}
\def\cii{c_2}
\def\ciii{c_3}

\def\gv{ g_V }
\def\fstate{\mathcal{F}}
\def\ph{|\vec{p_h}|}

\def\nn{\nonumber\\ }
\def\rd{{\rm d}}

\title{Higgs form factors in Associated Production }

\author[a,b]{Gino Isidori,}

\author[a]{and Michael Trott}

\affiliation[a]{Theory Division, Physics Department, CERN, CH-1211 Geneva 23, Switzerland}
\affiliation[b]{INFN, Laboratori Nazionali di Frascati, I-00044 Frascati, Italy}

\abstract{We further develop a form factor formalism
characterizing ano\-malous interactions of the Higgs-like boson ($h$)
to massive electroweak vector bosons ($V$) and generic bilinear fermion states ($\fstate$).
Employing this approach, we examine the sensitivity of $pp \to \fstate \rightarrow Vh$ associated production to physics beyond the 
Standard Model, and compare it to the corresponding sensitivity of $h \rightarrow V \fstate$ decays. 
We discuss how determining the $Vh$  invariant-mass distribution in associated production at LHC is a key ingredient  
for model-independent determinations of $h V \fstate$ interactions.
We also provide a general discussion about the power counting of the form factor's
momentum dependence in a generic effective field theory  approach, analyzing in particular 
how  effective theories  
based on a linear and non-linear realization of the $\rm SU(2)_L \times U(1)_Y$ gauge symmetry map into the form factor formalism.
We point out how measurements of the differential spectra characterizing $h \rightarrow V \fstate$ decays and $pp \to \fstate \rightarrow Vh$ associated production
could be the leading indication of the presence of a nonlinear realization of the $\rm SU(2)_L \times U(1)_Y$ gauge symmetry.} 

\begin{document} 
\maketitle
\flushbottom

\section{Introduction}

In this paper we study the utility of scalar production in association with a single massive electroweak vector boson
($W^\pm$ or $Z^0$, generically denoted by $V$) in constraining physics beyond the Standard Model (SM), and determining if the observed Higgs-like Boson is 
precisely the SM Higgs Boson. Further developing the formalism introduced in Ref.~\cite{Isidori:2013cla}, we characterize the  interactions of the Higgs-like boson, $h$, to $V$ and a generic bilinear 
fermion state $|\fstate\rangle = |\bar\psi \psi\rangle$ by means of a set of form factors, depending only on the $\fstate$
invariant mass ($q^2$).  This formalism has a twofold advantage.
On the one hand it is very general:  the information about physics beyond the SM that can be extracted from processes of the 
type  $h\to V \fstate$ (or $pp \to \fstate \to hV$) can be encoded in the $h V \fstate$ form factors, 
provided the creation (or annihilation) of the state $\fstate$ is described by a  local bilinear fermion current. 
On the other hand, the formalism allows us to relate different physical processes based on the same fundamental 
three-point $h V \fstate$  Green's function. In particular, it allows us to compare, under a minimal set of assumptions, 
the sensitivity to physics beyond the SM (BSM) of $\fstate \rightarrow Vh$ associated production to that of $h \rightarrow V \fstate$ decays. 

The production of $h V$ at large invariant mass is potentially accessible at the LHC and
offers a key probe of the $h V \fstate$ form factor behavior at large $q^2$. Probing for the large $q^2$ behavior of the
three-point $h V \fstate$  Green's function is not possible by looking at $h \rightarrow V \, F$ decays only~\cite{Isidori:2013cla}, which
are restricted to low $q^2$. We show how this 
information can be extracted from future LHC data by measuring the appropriate differential distributions and then
mapping it onto an underlying EFT formalism. 

The extra sensitivity to BSM effects in the reconstructed differential distributions follows from the introduction of non-standard $q^2$ dependence 
in the form factors. In this respect, our decomposition into $h V \fstate$ form factors  is a more detailed study than past analyses of the $h \rightarrow V V^*$ amplitude~\cite{Bolognesi:2012mm,Boughezal:2012tz,Gao:2010qx,DeRujula:2010ys}.
While these previous works have analyzed in generality the possible Lorentz-structure of the $h \rightarrow V V^*$ amplitude,
even  assuming non-standard spin and parity properties for $h$, they have frequently implicitly assumed that the $h \rightarrow V V^*$ amplitude
can be unambiguously determined by the structure of an (hypothetical) on-shell $h V V$ vertex,  or if this assumption was not made, they have practically neglected the $q^2$ dependence of the form factor in detailed analyses. Since the $h V V$ interaction is not kinematically 
accessible given the experimental values of $m_h$ and $V$, it is important to include the effects of non-standard 
$h V \fstate$ contact interactions.\footnote{
Previous studies on associated production 
of the Higgs-like Boson include Ref.~\cite{Glashow:1978ab,Ciccolini:2003jy,Stange:1994bb,Kleiss:1990vca,Belyaev:2008yj,Frederix:2011qg,Ferrera:2011bk,Ellis:2012xd,Ellis:2013ywa,Masso:2012eq,Dawson:2012gs,Englert:2013tya}. See also Ref.~\cite{Azatov:2012qz,Grojean:2013kd,Contino:2013kra,Artoisenet:2013jma,Alonso:2012pz} for some other recent  analyses in the EFT context.}

Generalizing the analysis of Ref.~\cite{Isidori:2013cla}, in this work we develop the  decomposition of the $h V \fstate$ amplitude
in the presence of generic contact interactions, 
 whose effect can be incorporated in the form-factor formalism via non-SM $q^2$ dependence.
We also analyze the power counting of the form factor's
momentum dependence in a generic effective field theory (EFT) approach. Given the experimental evidence of a Higgs-like 
boson, and the absence of non-SM particles at the LHC, 
two effective theories are of particular current interest: an EFT with a non-linear realization of $\rm SU(2)_L \times U(1)_Y$, 
where  $h$ is introduced as an $\rm SU(2)_L \times U(1)_Y$ scalar singlet, and the EFT based on the linear realization of
the SM gauge group, where $h$ belongs to an $\rm SU(2)_L$ doublet. We show how these two EFT's 
map into the  form factors. As a by-product, we clarify how possible deviations from the SM in the $h V \fstate$ 
differential spectra
could provide a clear clue to distinguish between these two EFT approaches. The mapping into an underlying
EFT is in fact required to evaluate the form factors from the theoretical point of view. However, 
this fact does not diminish the usefulness of the form factor formalism as an intermediate analysis step in a decomposition of the
$h\to V \fstate, \fstate \to hV$ amplitudes that allows us to potentially distinguish the two EFT approaches, as we show.

In other words, the form-factor approach provides us a compact and very general tool to map possible
experimental measurements into information on some underlying field theory, minimizing the assumptions on the latter
(for instance, without specifying  of how $\rm SU(2)_L \times U(1)_Y$ is realized in the scalar sector). 
Of course one must map onto an explicit EFT to perform perturbative corrections and define the amplitude beyond leading order. 
However, one of the purposes of this paper is to make clear the manifest utility, and the technical simplification,  that 
using form factors as an intermediate step in the calculation (and in data analysis) supplies.

The outline of the paper is as follows. In the next Section we introduce the $h V \fstate$ form factors and discuss how they modify 
of the SM cross section formulae for associated Higgs production. The general power counting decomposition of the 
the form factors is presented in Section~\ref{PowerCount}, and their explicit evaluation within the linear and nonlinear EFTs 
 is discussed in Section~\ref{EFTdefinition}. A numerical study of BSM effects in associated production is presented in 
Section~\ref{distributions}. The results are summarized in Section~\ref{conclude}.

\section{Amplitude Decomposition}\label{amps}
Following Ref.~\cite{Isidori:2013cla}, we start with the general decomposition of the $h\to V \fstate$ transition amplitude, 
where $V$ is an (approximately) on-shell massive weak gauge boson, 
and $\fstate$ is a state generated at tree level by the electroweak charged or neutral currents,  $J^V_\mu$, defined by 
\be
\label{currentdefn}
\mathcal{L}^{\rm SM}_{J} =  \frac{e }{ \sqrt{2}\, \sin \theta_W}  J_\mu^{\pm} W^\mu_{\pm}   + \frac{e}{\sin \theta_W \cos \theta_W}  J_\mu^0 Z^\mu
=\sum_{V}  C_V g_V J^V_\mu  V^\mu~.
\ee
Here  $\gv=\{g_2,g_2/\cos\theta_W\}$, $g_2=e/\sin\theta_W$, and $C_V = \{1/\sqrt{2},1\}$ 
are the coupling and normalization factors for $V=\{W^\pm,Z^0\}$. 
Defining $J^{\fstate_V}_\mu =\langle \fstate | J^V_\mu |0 \rangle$,  the transition amplitude
 can be decomposed in terms of four independent Lorentz structures
\bea
 \cA^{\fstate}_V &=&   \cA\!\left[ h \to V(\vareps,p) \fstate(q)\right ] = \frac{C_V \gv^2 m_V }{(q^2 - m_V^2)} ~ \vareps_\mu  J^{\fstate_V}_\nu T_{V}^{\mu\nu} \label{eq:gendec}~, \\
 T_{V}^{\mu\nu} &=&  \left[f^V_1({q}^2) g^{\mu\nu}  
 +  f^V_2({q}^2)  {q}^\mu {q}^\nu  
 +  f^V_3({q}^2)({p}\cdot {q}~g^{\mu\nu} -{q}^\mu {p}^\nu)  + f^V_4({q}^2) \epsilon^{\mu\nu\rho\sigma} {p}_\rho {q}_\sigma \right]~. \quad 
 \nonumber
\eea
The amplitude in Eq.~(\ref{eq:gendec}) controls both the $h \to V \fstate$ decay and the $Vh$ associated 
production process, $\fstate \rightarrow V h$. Denoting $q$ the total momentum of the initial state  in the partonic process $ \bar{\psi}\psi \rightarrow V h$,
we can write   
\be
\cA \left[  \bar{\psi} \psi(q) \to h \, V(\vareps,-p) \right ] = \cA^{ \psi\bar{\psi} }_V,
\ee
 where $\cA^{\bar{\psi}  \psi }_V$ is decomposed exactly as in Eq.~(\ref{eq:gendec}). 
The important difference between $h$ decays and $Vh$ associated production is the allowed kinematical 
region probed by the process:  $ 0< q^2 < (m_h -m_V)^2$ in $h$ decays, and
$q^2 > (m_h +m_V)^2$ in the $Vh$ associated production. This fact can lead to an enhanced sensitivity to BSM effects in kinematic distributions
of associated production compared to kinematic distributions of $h\to V \fstate$ decays.
  
The partonic cross sections for the $Vh$ associated production within the SM, at fixed $q^2$,  are well known \cite{Ellis:1975ap}
\bea
\sigma ( \bar{\psi} \psi \rightarrow Z h)^{\rm SM}  &=& \sigma_0^{Zh} \, \frac{|\vec{p}_h|}{\sqrt{q^2}} \, \frac{\ph^2 + 3 m_Z^2}{(q^2- m_Z^2)^2}~,  
\qquad \sigma_0^{Zh} = \frac{2 \, \pi \, \alpha^2 [(g^{L}_\psi)^2 + (g^{R}_\psi)^2]}{12 \, N_c \, \sin^4 \theta_W \, \cos^4 \theta_W},
\nonumber \\
\sigma ( \bar{\psi}_j \psi_i \rightarrow W h)^{\rm SM} &=& \sigma_{0, ij}^{WH}  \, \frac{\ph}{\sqrt{q^2}} \, \frac{\ph^2 + 3 m_W^2}{(q^2- m_W^2)^2}~,
\quad 
\quad \quad \sigma_{0, ij}^{Wh} = \frac{\pi \, \alpha^2 |V_{ij}|^2}{18 \,  \sin^4 \theta_W}~,
\eea
where $\ph = [(q^4 + m_h^4 + m_V^4 - 2 q^2 m_h^2 - 2q^2 m_V^2 -2 m_V^2 m_h^2)/(4 q^2)]^{1/2}$
is the center of mass momentum of the Higgs-like boson and $V_{ij}$ denotes CKM matrix elements.\footnote{In this paper we will use $i,j$ for flavour indices.} Here  
\be
g^L_\psi  = 2 \, T^\psi_3 - 2 \, Q_\psi  \, \sin^2 \theta_W~, \qquad g^R_\psi = - 2 \, Q_\psi \, \sin^2 \theta_W~,
\ee
are the left-handed and right-handed coupling of the $Z$ boson, normalized as in Eq.~(\ref{currentdefn}).
Using the SM rate as a normalization, the generalized partonic results in the case of the form factor decomposition 
 in Eq.~(\ref{eq:gendec}) assumes the following simple form
\bea
\frac{\sigma^{\rm BSM} (\psi \, \bar{\psi} \rightarrow V h)}{\sigma^{\rm SM} (\psi \, \bar{\psi} \rightarrow V h)} &\equiv & R_V(q^2) =  \left| f^V_1 (q^2) \right|^2  
+  3 ~{\rm Re} \left[f^V_1(q^2)  f^{V*}_3(q^2)\right]  \frac{ m_V^2 (q^2 + m_V^2 - m_h^2)}{ \ph^2 +3 m_V^2 } \nonumber \\
&& + \frac{ m_V^2 q^2}{ \ph^2 +3 m_V^2 } \left[ \left| f^V_3(q^2) \right|^2  (3  m_V^2 + 2  \ph^2) + 2  \ph^2  \left| f^V_4(q^2)\right| \right].
\label{eq:Rcross}
\eea

\subsection{Generalization to arbitrary currents}
\label{sec:gencurr}
In writing Eq.~(\ref{eq:gendec}) we have assumed that the state $\fstate$ is generated only by the weak currents in Eq.~(\ref{currentdefn}).
This assumption is motivated by the fact that this is the only allowed tree-level contribution to the $h\to V \fstate$ amplitude within the SM.
 With this assumption, the decomposition
in Eq.~(\ref{currentdefn}) is aimed at characterizing anomalous Higgs interactions with the ${\rm SU(2)_L \times U(1)_Y}$ gauge bosons.

Nevertheless, this assumption is not necessarily a good approximation in generic extensions of the SM.
The decomposition in Eq.~(\ref{eq:gendec}) can be easily generalized to all cases where 
$\fstate$ is generated by generic local left-handed or right-handed fermion current,
\be
(J^{ij}_L)_\mu = \bar{\psi}^i_L \gamma_\mu  \psi^j_L \qquad {\rm or} \qquad  (J^{ij}_R)_\mu = \bar{\psi}^i_R \gamma_\mu \psi^j_R~,
\ee
provided we enlarge the number of independent form factors. In general, for each independent current $J$ 
contributing to the final state $\fstate$,  we have a set of four independent form factors ($f_{1-4}^{V_J}$) associated with
the four different Lorentz structures appearing in Eq.~(\ref{eq:gendec}).
In the limit of a $\rm SU(3)^5$ unbroken flavour symmetry, there are five independent currents 
$(J^f)_\mu=\bar f\gamma_\mu f$, where $f = 
\{u_R,d_R,e_R, L_L,Q_L\}$, and at most two of them ($J^\fstate_L,~J^\fstate_R$) can create or annihilate the state $\fstate$ at the tree level.
The trivial generalization of Eq.~(\ref{eq:gendec}) then reads
\bea
 \cA^{\fstate}_V &=&  \frac{C_V \gv^2 m_V  }{(q^2 - m_V^2)} \sum_{J=J^\fstate_L, J^\fstate_R} 
   \vareps_\mu ~\langle \fstate | J_\nu |0 \rangle~  T_{V_J}^{\mu\nu}~, 
   \label{eq:gendecLR} 
\eea
where  $T_{V_J}^{\mu\nu}$  contains the form factor set  $f_{i}^{V_J}$. Using this generalization is particularly simple 
for inclusive observables obtained integrating over angular variables relative to the final states $\fstate$, such as the cross-Section in Eq.~(\ref{eq:Rcross}). In this case, the left-handed and right-handed currents do not interfere. It is also useful to distinguish between 
charged and neutral currents:
\begin{description}
\item[\em Charged currents.] Since the tree-level SM amplitude is purely left-handed, non-SM contributions due to 
right-handed charged currents do not interfere with the leading SM amplitude and can safely be neglected.
\item[\em Neutral currents.] In this case the result in  Eq.~(\ref{eq:Rcross}) can be generalized to 
\be
\frac{\sigma^{\rm BSM} (\psi \, \bar{\psi} \rightarrow V h)}{\sigma^{\rm SM} (\psi \, \bar{\psi} \rightarrow V h)} = \frac{(g^{L}_\psi)^2 } { (g^{L}_\psi)^2 + (g^{R}_\psi)^2}~R_L(q^2) + \frac{(g^{R}_\psi)^2 } { (g^{L}_\psi)^2 + (g^{R}_\psi)^2} ~R_R(q^2)~, 
\ee
with $R_{L}$ ($R_R$) identical to $R_V$ in Eq.~(\ref{eq:Rcross}) but containing only the 
left-handed (right-handed) currents, and corresponding form factors.
\end{description}

\section{General EFT power counting for the form factors}\label{PowerCount}

We can determine the power counting for the $q^2$ dependence of the different form factors in general.
This is of interest, as the BSM momentum dependence present in the deformation of the amplitude from the SM case is the origin of the enhanced sensitivity to BSM effects 
in some kinematic distributions compared to the total integrated rate. This point was recently explored in some detail in 
Refs.~\cite{Isidori:2013cla,Grinstein:2013vsa} for the case of $h\to V \fstate$ decays. 

For simplicity in this discussion, we focus on the case where  
the Higgs boson belongs to an $\rm SU(2)_L$ doublet -- in other words, to the case of 
a linear realization of the $\rm SU(2)_L \times U(1)_Y$ gauge symmetry.
The differences and similarities with respect to the non-linear realization of $\rm SU(2)_L \times U(1)_Y$
will be discussed at the end of this Section and in Sec.~\ref{EFTdefinition}.
We also limit ourself to evaluating the power counting associated with 
the dimensionality of the operators in the EFT expansion, ignoring the counting associated with powers of the 
gauge couplings. Our purpose here is to discuss some general properties of the form-factor expansion that just follow from the field content and Lorentz symmetry of the underlying theory.

\subsection{General decomposition}

As a first step, we expand each  form factor  around $q^2=m^2_V$ in the following way,
\be
\label{eq:genffexp}
 f^{V_J}_i({q}^2) =   f^{V(0)}_i  +  \sum_{n\geq 1} f^{V_J(n)}_i \left(\frac{q^2-m_V^2}{m_V^2}\right)^n  +  
  \frac{q^2-m_V^2}{m_V^2}  \Delta f^{V_J(\rm LD)}_i({q}^2)~,
\ee
separating the {\em constant terms}, $f^{V(0)}_i$, the {\em slopes}, $f^{V_J(n\geq 1)}_i$,  and the {\em long distance} 
contribution due to the propagation of real intermediate states, $\Delta f^{V_J(\rm LD)}_i({q}^2)$. 
We use the label $V_J$ ($V$) to denote  the form factors components that 
can (cannot) depend on the specific choice of the current. 

By construction,  the constant terms are associated with the 
$q^2 \to m^2_V$ pole of the $hV\fstate$  amplitude.  As such, they can be put in one-to-one correspondence 
with the general decomposition of a hypothetical on-shell $hVV$ amplitude (that is not kinematically accessible 
given the experimental values of $m_h$ and $m_V$),  and select only the $J^V$ component
of the current.
The relation $f^{V}_2(m_V^2)=-f^{V}_1(m_V^2)/m_V^2$
  implies 
$f^{V(0)}_2=-f^{V(0)}_1/m_V^2$ hence there are only three independent constant terms for each~$V$ \cite{Isidori:2013cla}. 

With a suitable choice of operator basis, the $f^{V_J(n\geq1)}_i$  can
always be associated with a $hV\fstate$ contact interaction.
For this reason, the form factor  slopes are dependent on the structure of the current creating the state $\fstate$.  
There are four independent $f^{V_J(n\geq 1)}_i$ for each $J$ and $n$. 

Finally,  $\Delta f^{V_J(\rm LD)}_i({q}^2)$  denote the sum of singular terms related to 
light poles in the amplitude. These poles are due to the electromagnetic current, for the $q^2\to0$ pole,
or the $q^2\to m^2_{\rm had}$ poles of 
single hadronic states with mass $m_{\rm had}$ and the same quantum number of $\fstate$. 
Note that there are other non-analytic terms due to multi-particle cuts.

\subsection{Constant terms}

To determine the power counting of the constant terms we start from the general decomposition of the $hV(\vareps,p)V(\eps,q)$ amplitude:
\be
 \cA_{VV}  \propto     \vareps_\mu  \epsilon_\nu   \left[f^{V(0)}_1  g^{\mu\nu}  
 +  f^{V(0)}_3 ({p}\cdot {q}~g^{\mu\nu} -{q}^\mu {p}^\nu)  + f^{V(0)}_4  \epsilon^{\mu\nu\rho\sigma} {p}_\rho {q}_\sigma \right] ~.
 \ee
When considering the effective operators generating  $\cA_{VV}$ in a linear realization, they are constructed combining Higgs fields, $H \sim O(p)$, gauge stress tensors, $F^{\mu\nu}
\sim O(p^2)$,
and an arbitrary number of covariant derivatives, $D_\mu \sim O(p)$.  $\rm SU(2)_L$ invariance for this amplitude forbids terms with odd powers of $H$, and we need at least 
a $H H^\dagger$ pair to extract the physical Higgs-boson field $h$.  Given this structure, $f^{V(0)}_1$ can be non-vanishing at $O(p^4)$
while $f^{V(0)}_{3,4}$ can be non-vanishing only at $O(p^6)$. Since within this EFT the only relevant  $O(p^4)$ operator is the SM Higgs kinetic term, 
we obtain

\be
f^{V(0)}_{1} - f^{V(0){\rm SM} }_1  = O(p^6)~, \qquad  f^{V(0)}_{3,4}    = O(p^6)~.  
\ee
Here and in the following the notation  $X=O(p^n)$ stands for: $X$ is non-vanishing at $O(p^n)$, irrespective 
of the dimension of $X$.

\subsection{Slopes in $q^2$ from contact interactions}

Contact interactions with a generic current $J_\mu$ generates an amplitude of the type
\bea
 \cA^{\fstate_J}_V  &\propto&  \vareps_\mu  J^{\fstate_J}_\nu    \left[ f_1^{V_J(1)}  g^{\mu\nu}  
 +  f^{V_J(1)}_2   {q}^\mu {q}^\nu 
 +  f^{V_J(1)}_3  ({p}\cdot {q}~g^{\mu\nu} -{q}^\mu {p}^\nu)  + f^{V_J(1)}_4  \epsilon^{\mu\nu\rho\sigma} {p}_\rho {q}_\sigma \right]~,
  \nonumber \\
&&  + O(q^2 -m_V^2)~.
\eea
The currents in Eq.~(\ref{eq:gendecLR}) are $O(p^3)$ and $\rm SU(2)_L$ invariant.  
The $\rm SU(2)_L$ invariance of the amplitude then implies at least two Higgs fields, and the counting of derivatives and/or gauge fields, 
both of $O(p)$, implies that $f^{V_J(1)}_1$ can be generated at $O(p^6)$, while $f^{V_J(1)}_{2,3,4}$ can appear only at $O(p^8)$.
The generalization to higher slopes can easily be obtained noting that the extra $q^2$ dependence implies (at least) two more derivatives:
\be
f^{V_J(n \geq 1)}_{1}  = O(p^{4+2n})~, \qquad  f^{V_J(n \geq 1)}_{2,3,4}    = O(p^{6+2n})~.  
\ee

The  decomposition (\ref{eq:gendecLR}) allow us to consider also  scalar currents  of the type $\bar \psi_{L(R)} \psi_{R(L)}$, with a
covariant derivative acting on one (or both) fermion fields.   In this case the power counting is the same as above:  $\rm SU(2)_L$ invariance 
implies an odd power of Higgs fields, but the structure of the current is such that $J^{\fstate_J}_\nu$ is at least of $O(p^4)$ with a covariant  derivative acting on the fermion fields. By construction, 
our decomposition of the amplitude does not allow us to include local operators with a generic tensor 
current ($\bar \psi_{L(R)} \sigma^{\mu\nu} \psi_{R(L)}$).\footnote{In general,  contributions due to tensor currents do not need not be loop suppressed~\cite{Jenkins:2013fya}. However, their inclusion require an extension of the present formalism that is beyond the scope 
of the present work.}

\subsection{The photon pole}

In most realistic BSM constructions, the only relevant additional pole in the amplitude is 
the one occurring at $q^2=0$ due the  photon propagator, for states $\fstate$ 
 such that $\langle \fstate  | J^{\rm em}_\mu |0 \rangle \not=0$, where $J^{\rm em}_\mu$ is the electromagnetic current.
In this limit, the structure of $\Delta f^{V_J(\rm LD)}_i({q}^2)$  
can be put in one-to-one correspondence to the general decomposition of the on-shell $hZ\gamma$ amplitude
($q^2$ dependent terms that cancel the pole at $q^2=0$ can be absorbed in the form factor slopes).
The latter can be decomposed in terms of  two independent Lorentz structures,
\be
 \cA_{Z\gamma}  \propto     \vareps_\mu  \epsilon_\nu   \left[ ~g_3~({p}\cdot {q}~g^{\mu\nu} -{q}^\mu {p}^\nu)  +  g_4 ~\epsilon^{\mu\nu\rho\sigma} {p}_\rho {q}_\sigma \right] ~.
 \ee
Within the SM the effective couplings $g_{3,4}$ are different from zero only beyond tree level. 
This fact does not imply that such effects must be loop suppressed in a BSM sector.
Within the EFT approach,  using the same power counting adopted for the  $\cA_{VV}$ amplitude, it follows that $g_{3,4}$ can be different from zero 
starting  at $O(p^6)$. The resulting  $\Delta f^{Z_J(\rm LD)}_i({q}^2)$  does depend on the nature of  $\fstate$; however, the dependence is
 fully specified by the electric charge $Q_{\psi}$ of the fermions in $| \fstate  \rangle = | \bar \psi \psi \rangle$:
 \be
\Delta f^{Z_J(\rm LD)}_{3,4}({q}^2)  \propto g_{3,4} \frac{Q_{\psi}}{q^2} = O(p^{6})~,  \qquad 
\Delta f^{Z_J(\rm LD)}_{1,2}({q}^2)  =0~.
\ee

\subsection{Summary}
Due to the above discussion, the power counting within the linear EFT approach implies the following general expansion of the 
form factors up to $O(p^6)$:
\bea
f^{V_J(0)}_{1} - f^{V_J(0){\rm SM} }_1  &=&  a^{(1)}_{V}  \frac{v^2}{\Lambda^2} + b^{(1)}_{V_J}  \frac{q^2 - m_V^2}{\Lambda^2}~, 
\label{eq:zero} \\
f^{V(0)}_{i} - f^{V(0){\rm SM} }_{i}  &=&  a^{(i)}_{V}  \frac{1}{\Lambda^2}~,  \qquad (i=2,3,4)~, \\
\Delta f^{Z_J(\rm LD)}_{3,4}  - \Delta f^{Z_J(\rm LD){\rm SM}}_{3,4} 
 &=&  c^{(i)}_{Z} Q_{\psi} \frac{v^2}{\Lambda^2} \frac{1}{ q^2}~,  \qquad (i=3,4)~,
 \label{eq:genPC}
\eea
where the $a^{(1)}_{V}$,  $b^{(1)}_{V_J}$, and $c^{(i)}_{Z}$ are $O(1)$ couplings. 

As we will explicitly  illustrate in the next Section, a very similar 
expansion is recovered also in the non-linear case. This is somewhat expected. Both theories are defined by derivative expansions with the same field content and the appearance of the momentum dependence is constrained by Lorentz invariance. As such, the only form factor that can have $q^2$ dependence at sub-leading order is $f^{V(0)}_1$.
In this manner, although the linear realization was used
in this discussion, the structure of the form factors that has been identified is more general in many respects.
This is a reason why a general analysis in terms of a form factor decomposition is a useful intermediate step
until the most suitable underlying EFT is experimentally clarified.

The form factor formalism
can also accommodate a non-SM momentum dependence (consistent with crossing symmetry and Lorentz invariance) that differs from the polynomial expansion around $m_V^2$
of the $f_i$ that we have assumed. The assumed dependence is suitable for a local EFT expansion of the form factors.  Non-SM momentum dependence, with unsuppressed coefficients will generally lead to unitarity violation that
is proximate to the scale $\sim v$. We restrict our attention to a polynomial derivative expansion of the form factors
in what follows, as appropriate for local EFTs with no additional light states.  
We also generally consider suppressed coefficients in the derivative expansion of the momentum dependence of the
form factors for this reason.  See the discussion in Section \ref{spectra} for more details on this point.

As far as the decomposition of the form factors is concerned, a key difference between the two EFT approaches
concerned the leading term, $f^{ V(0)}_{1} = -m_V^2 f^{ V(0)}_{2}$, i.e.~the only  term that does not vanish at the tree level within the SM. 
Within the linear EFT approach, the only leading operator contributing 
to the $hV\fstate$ amplitude at the tree level is the SM Higgs kinetic term. 
This fact is responsible for the absence of $O(1)$ corrections in $f^{V(0)}_{1}$.  
Within the non-linear case, the SM is recovered only with an explicit tuning of the 
leading operators. In the absence of such tuning,  $f^{ V(0)}_{1}$  can receive  $O(1)$ corrections
in the non-linear EFT approach compared to the SM expectation.

\section{The form factors in explicit EFTs}\label{EFTdefinition}
We now decompose the form factors in Eq.~(\ref{eq:genffexp}) in terms of local operators in an Effective Lagrangian. 
This is a necessary step to calculate the amplitude beyond tree level in a BSM scenario. 
However, as we have emphasized, there is more than one EFT to choose from. In the following, we provide an illustrative decomposition of the form factors in 
two EFTs of particular current interest: an EFT with a non-linear realization of $\rm SU(2)_L \times U(1)_Y$, 
where the massive Higgs-like boson $h$ is introduced as an $\rm SU(2)_L \times U(1)_Y$ scalar singlet~\cite{Grinstein:2007iv,Contino:2010mh}
(see also Ref.~\cite{Burgess:1999ha}), 
and the linear realization EFT where $h$ belongs to an $\rm SU(2)_L$ doublet $H$. As already emphasized, 
these two EFT approachs are in fact  distinct. We comment about the differences that arise  in the specific case of the $hV\fstate$ form factors 
at the end of this Section.

In the following discussion, our aim is not to provide a systematic evaluation of the 
form factors in these two theories including all terms in the sub-leading operator basis. In both cases, the number of sub-leading operators
is sizable, and such an analysis is not particularly illuminating. 
Instead, we restrict our attention to contributions to the physical process due to anomalous Higgs-like Boson interactions that
are illustrative of non-SM $q^2$ dependence in the form factors.
In particular, we neglect local interactions with currents different from $J^V$,
and possible contributions due to a modified $hZ\gamma$ amplitude.

\subsection{Non-Linear realization of $\rm SU(2)_L \times U(1)_Y$}
The non-linear EFT is based on the assumption of a global symmetry $\rm \cG=SU(2)_L \times SU(2)_R$ existing in a BSM sector,
that is spontaneously broken into the diagonal subgroup 
$\rm SU(2)_{L+R}$ (corresponding to custodial symmetry). The Goldstone bosons resulting from this spontaneous symmetry breaking, 
denoted by $\pi^a$,  are described by the field 
\be
\Sigma(x) = e^{i \sigma_a \, \pi^a/v}  \qquad a = 1,2,3~,
\ee
that  transforms (linearly) as a $(2,\bar 2)$ of $\cG$.
The $\rm SU(2)_L \times U(1)_Y$ subgroup of $\rm SU(2)_L \times SU(2)_R$ is weakly gauged, such that the $\pi^a$ can be identified 
with the longitudinal components of the SM massive gauge bosons. The SM gauge symmetry is nonlinearly realized. An $h$ field is introduced as a massive $0^+$ scalar, 
which is a singlet under $\cG$. The leading order terms involving the observed scalar field in the derivative expansion of this EFT are
\bea
\cL^{(2)}_{EWh} &=& \mathcal{L}_h  + \cO_{LO} + \cO_Y,   \\ 
\mathcal{L}_h  &=&   \frac{1}{2} (\partial_\mu h)^2 -  \frac{1}{2} \, m_h^2 \, h^2  \left[ 1  +  d_3 \left(\frac{h}{v}\right) 
    + \frac{d_4}{4} \left(\frac{h}{v}\right)^2 + \cdots   \right] \\
    \cO_{LO} &=& \frac{v^2}{4}  \left[ 1 + 2 \ci  \left(\frac{h}{v}\right)  +\cdots \right] {\rm Tr} \left[ (D_\mu \Sigma)^\dagger \, D^\mu \Sigma\right].   
\eea
Here $\cO_Y$ are the Yukawa interactions. The coefficients have been normalized to define the 
suppression scale in the expansion in powers of $h$ as $1/v$. 
The most famous UV completion of this theory is the singular case where the coefficients
in the $(h/v)$ expansion are fixed such that  $h$ and the Goldstone bosons can be  
combined into a single linear multiplet  $H = (1 + h/v) \Sigma$. In this case $\cL^{(2)}_{EWh}$ coincides with the 
SM Higgs boson Lagrangian. For the purpose of this work, the unknown coefficients present in $\cL^{(2)}_{EWh}$
are arbitrary numerical parameters subject to experimental constraints: present data suggests their values are proximate to the SM limit, but we are far from having ruled out a more general  framework. The presence of a light scalar in the spectrum can postpone the usual unitarity problems of massive vector bosons to
significantly higher scales. Considering LHC's discovery reach, a true "Higgs-like Boson", proximate in its properties to the SM Higgs boson, and described by this theory, is a consistent EFT framework.

The complete list of local operators to the next order in the derivative expansion, under the hypothesis of Minimal Flavour Violation~\cite{Chivukula:1987py,Hall:1990ac,D'Ambrosio:2002ex,Cirigliano:2005ck}, can be found in Ref.~\cite{Alonso:2012px,Buchalla:2013rka}. Here we restrict the attention to operators that are CP even and do not contain any source of custodial symmetry breaking\footnote{More precisely, we neglect operators that either have a $\sigma_3$ in a trace, are proportional to the hypercharge coupling ($g_1$), or are related to SM fermion masses through the equations of motion}. In this limit, only two $O(p^4)$ operators lead to a  non-vanishing $hVV^*$ amplitude at the tree level:
\bea
 \mathcal{L}_{EWh}^{(4)} &=&  \sum_{i = W, W\partial h, \ldots}  c_i^{\rm (4)} \mathcal{O}_i \\
\mathcal{O}_W &=&  g_2 \, D_\mu W_a^{\mu \, \nu} {\rm Tr} (\Sigma^\dagger \, i \sigma^a  \overleftrightarrow{D}_\nu \Sigma) F_W~,  \\
\mathcal{O}_{W \partial h} &=& g_2 \, W_a^{\mu \, \nu}  {\rm Tr} (\Sigma^\dagger \, i \sigma^a \,  \overleftrightarrow{D}_\mu \Sigma) \partial_\nu F_{W \partial H}~. 
\eea
Here the $F_i$ are generic polynomials in $(h/v)^n$ and we have used the notation
\bea
D_\mu \, \Sigma &=& \partial_\mu \, \Sigma + \frac{i \, g_2}{2} \, W^a_\mu \, \sigma_a \, \Sigma - \frac{i \, g_1}{2} \, B_\mu \, \Sigma \, \sigma_3, \nn
\Sigma^\dagger \, \sigma^a \overleftrightarrow{D}_\nu \Sigma &=& \Sigma^\dagger \, \sigma^a \, (D_\nu \Sigma) - (D_\nu \Sigma)^\dagger \, \sigma^a \, \Sigma.
\eea
Considering also the leading order operator ($\cO_{LO}$), we have three operators generating 
a  non-vanishing $hVV^*$ amplitude at the tree level. After expanding the $(h/v)^n$ polynomials 
to the first order,  to be consistent with our past work \cite{Isidori:2013cla} we define  the Wilson coefficients $c_i$ 
for the $h$-linearized terms,
\bea
\hat \cO_{LO} &=& \frac{v  \, \ci}{2}\  h\, {\rm Tr} \left[ (D_\mu \Sigma)^\dagger \, D^\mu \Sigma\right],  \nn
\hat \cO_W &=&  \frac{g_2  \, \cii}{ v}\ h \, D_\mu W_a^{\mu  \nu} {\rm Tr} \left[\Sigma^\dagger \, i \tau^a  \overleftrightarrow{D}_\nu \Sigma\right],  \nn
\hat O_{W\! \partial H} &=& \frac{g_2  \, \, \ciii  }{ v}\ (\partial_\nu h) \, W_a^{\mu \nu}  {\rm Tr} \left[ \Sigma^\dagger \, i \tau^a \,  \overleftrightarrow{D}_\mu \Sigma \right]\,, \label{eq:Op}
\eea
such that the SM case correspond to $c_i^{\rm SM} = (1,0,0)$~.
The form factor basis expanded in terms
of these operators is  
\bea
f_1^V({q}^2)  &=&   \ci   +  g_2^2  \, (\cii \, +\ciii) \, \left(1 + \frac{{q}^2}{m_V^2} \right)~, \quad \quad f_2^V({q}^2)  = - \frac{1}{m_V^2} \, \left[\ci  + 2 \, g_2^2  \, (\cii \, + \ciii) \right]~,  \nn
f_3^V({q}^2) &=& \frac{2 \, g_2^2 }{m_V^2}\ciii~,  \quad \quad \quad \quad \quad \quad \quad \quad  \quad \quad \quad \quad f_4^V({q}^2) = 0~. 
\label{eq:fNL}
\eea
As can be  seen, this result is consistent with the general decomposition in  Eq.~(\ref{eq:genPC}), provided 
we assume $c_{2,3}$ and $(c_{1}-1)$ to be $O(v^2/\Lambda^2)$, as expected by naive dimensional analysis.
A detailed analysis of the naive power counting in this EFT 
has recently been presented in Refs.~\cite{Buchalla:2013rka,Buchalla:2013eza}, where it has been shown 
that $(c_{1} -1)$ could be of $O(1)$ without spoiling the consistency of the loop expansion. 

Beside naive dimensional analysis, the expectation for the size of these parameters
must balance several dynamical considerations. First of all, we stress that in a nonlinear chiral EFT 
the cut off scale can be lower than its naive value, $\Lambda \sim 4\pi v$~\cite{Manohar:1983md}.
In the presence of strong interactions with a light $0^+$ state, large flavour, colour or Goldstone boson groups of dimension $\sim N$ 
in the underlying theory  can lead to a cut off scale of $\Lambda  \sim 4 \, \pi v /\sqrt{N}$ \cite{Soldate:1989fh,Chivukula:1992nw,Georgi:1992dw}. 

One can form more specific expectations for the unknown parameters $\left(c_1,c_2,c_3\right)$ at the cost of introducing
model dependence and moving outside of a totally general EFT framework.
For example, in the case of a composite Higgs (and the particular benchmark models discussed in ef.~\cite{Contino:2013kra})
one expects $(c_{1}-1)=O(v^2/f^2)$, while $c_{2,3}=O(v^2/M^2)$.
Here the parameters $f$ and $M$ are related by the unknown coupling constant $g_\star$
of the $h$ field to the new strong states. The parameters are related by $g_\star/M \simeq 1/f$ and one expects $1 \leq g_\star \leq 4 \, \pi$.
Further, the mass of the lightest
spin-1 resonances in the strongly interacting sector scale with $M$, although the exact masses also depend on further unknown couplings.
The exchange of such resonances can induce the operators with the coefficients $c_{2,3}$.
The experimental constraints on $f$ and $M$ are model dependent, but have strengthened significantly after run one at LHC.
Estimates of the bounds on these parameters \cite{Contino:2013kra} imply $v^2/f^2 \lesssim 0.1$ in these benchmark scenarios.
The size of the deviations shown in Fig.~\ref{fig:spectra3} can be compared to this benchmark expectation.
 
Adopting a bottom-up EFT approach allows us not to be overly concerned with these model considerations for the sake of a phenomenological analysis. 
However, we will generally only consider in
detail the range ${O}(10^{-2}-10^{-1})$, that we consider a rather natural choice for these parameters;  the parameters are chosen in the range of $0.01$ to $0.2$
for most of the figures shown in Fig.~\ref{fig:spectra3}. 
We also make a few illustrative plots for ${O}(0.1-1)$ values of the coefficients, in particular in the case labeled suppressed SM couplings
in Fig.~\ref{fig:spectra3} which we argue is likely already excluded by the total signal strength measurements
of ATLAS and CMS.

\subsection{Linear realization of $\rm SU(2)_L \times U(1)_Y$}
\label{eq:linEFT}

In the case of the linear EFT, the lowest-order Lagrangian coincides with the SM Lagrangian and the Higgs mechanism is at work to generate the masses of the 
observed (non scalar) SM particles. The complete (yet minimal) set of operators appearing at dimension-six (i.e.~the first non-trivial order
in the EFT expansion for higher dimensional operators contributing to processes conserving total lepton number) has been presented in 
Ref.~\cite{Grzadkowski:2010es}. Following the notation of Ref.~\cite{Grzadkowski:2010es},
the two operators modifying the $hVV^*$ amplitude at tree level, in the limit of unbroken CP  and  custodial symmetries, are 
\bea
\mathcal{P}_{WW} &=&  \frac{g_2^2}{ \Lambda^2} \, H^\dagger \,  H \, W^a_{\mu\, \nu} W^{a\,\mu \, \nu}, \quad \quad \mathcal{P}_{\Box} =  \frac{2}{\Lambda^2} \, (H^\dagger \, H) \, \Box \, (H^\dagger  \, H)~,
\label{ops}
\eea
and these operators lead to the decomposition 
\bea\label{localop}
f_1^V({q}^2)  &=& 1+ \frac{v^2 \, c_{\Box}}{\Lambda^2} ~,  \hspace{2.5cm}  f_2^V({q}^2)  =   - \frac{1}{m_V^2} \, \left(1+ \frac{v^2 \, c_{\Box}}{\Lambda^2} \right),  \nn
f_3^V({q}^2) &=& \frac{g_2^2 }{m_V^2} \left(\frac{v^2 \, c^V_{WW}}{\Lambda^2}\right)~,  \hspace{1.4cm} f_4^V({q}^2) = 0~. 
\eea
By construction, the $c_i$ are  dimension-less  couplings expected to be $O(1)$.
Given the definition of the operator $\mathcal{P}_{WW}$, an explicit  $V$ dependence does appear in its contribution to $f_3^V$: 
$c^V_{WW} = \{c_{WW}, \cos^2  \theta_W \, c_{WW}\}$. However, unbroken custodial symmetry is consistently recovered in the 
limit $g_1 \to 0$. 

In the operator basis of Ref.~\cite{Grzadkowski:2010es}, the momentum dependence in the form factors is generated by 
the contact operators,
\bea
\mathcal{P}_{HJ} &=&  \frac{4}{ \Lambda^2} \,  \left(H^\dagger i \, \overleftrightarrow{D}^I_\mu H\right) J^\mu~, \\
J^\mu &=&  \{ \bar{Q}_L \sigma_I \, \gamma^\mu Q_L~, \quad   \bar{Q}_L \sigma_3 \, \gamma^\mu Q_L~, \quad  \bar{u}_R  \gamma^\mu u_R~, \quad \bar{d}_R  \gamma^\mu d_R~, \nonumber \\
&&  \bar{L}_L \sigma_I \, \gamma^\mu L_L~, \quad \bar{L}_L \sigma_3 \, \gamma^\mu L_L~, \quad \bar{e}_R  \gamma^\mu e_R~\}. 
\eea
With a proper choice of couplings, such that the effective charged and neutral current combinations coincides with $J_V^\mu$ in 
Eq.~(\ref{currentdefn}), one generates a momentum dependence in $f^V_3$, with correlated constant shift in $f^V_2$, as in 
Eq.~(\ref{eq:fNL}).

\subsection{Linear vs. Nonlinear realization of $\rm SU(2)_L \times U(1)_Y$}\label{differences}

Despite the similarities of the decomposition of the form factors into the linear and non-linear EFT frameworks,
it is important to realize that,  at any fixed order in the EFT expansion, these theories are in fact distinct EFTs\footnote{A general discussion about the similarities and differences of these two EFT will be presented elsewhere.}.
This is perhaps counterintuitive, given the two theories have exactly the same gauge symmetry and the same field content. Which might naively be assumed to be sufficient to establish the equivalence of the two theories.
One should also note that a comparison of the number of free parameters (at a fixed order in the EFT expansion) contributing to a subset of processes, while ignoring correlated constraints 
due to the different (global) symmetry structure in the two theories, is also insufficient to establish the equivalence of the two approaches.
One way to see this is as follows.

Consider the analysis of $h\to V \ell^+\ell^-$ decays of Refs.~\cite{Isidori:2013cla,Grinstein:2013vsa}, where it was emphasized that 
studies of the $d \Gamma / d q^2$ spectrum are a sensitive probe of BSM effects compared to the total rate.
As discussed in Section~\ref{eq:linEFT}, within the linear EFT the non-standard $q^2$ dependence in the form factors 
is generated from contact operators of the type $\mathcal{P}_{HJ}$.  These operators contribute also to the 
(on-shell) $Z \rightarrow \ell^+ \, \ell^-$  decay when both Higgs doublets get a vev. 
As such, their Wilson coefficients are constrained by the electroweak precision tests performed at LEP.
This implies that in the linear EFT, the effects of  anomalous Higgs couplings present in the $d \Gamma / d q^2$ spectrum is correspondingly constrained by LEP measurements~\cite{pomarol}.

Now consider the nonlinear EFT. In this case the BSM momentum dependence in $h\to V \ell^+\ell^-$ 
is {\em not} directly related to deviations in $Z \rightarrow \ell^+ \, \ell^-$ 
as the scalar is not embedded in an $\rm SU(2)_L$ doublet. In other words, in this EFT the contact operators with  different powers of
 the $h$ field are unrelated in general. As a result, studies of the $d \Gamma / d q^2$ spectrum are in fact (fairly unique) probes of
Green's function with one external $h$ field, and this spectrum is not  as directly constrained by existing electroweak precision measurements of
$Z \rightarrow \ell^+ \, \ell^-$  decay.
Considering correlated constraints, the range of allowed parameters in the 
linear realization is a subset of the parameters that are still experimentally allowed in the non linear EFT.

\section{Associated Production as a probe of anomalous couplings}\label{distributions}

In this Section we consider the constraints that current and future associated production data can place 
on the parameters in the EFTs. As a starting hypothesis, we assume that by means of appropriate cuts (and 
background subtraction), the LHC experiments will be able to isolate and measure the 
$\sigma(q\bar q  \to Vh)$ s-channel cross section. 
Our main aim is to illustrate the utility of reporting the measured associated 
production spectrum as a function of the reconstructed $Vh$  $q^2$ distribution for constraining the underlying form factors, and hence the EFTs. 

\subsection{Current data}

Current associated production data reported by ATLAS \cite{ATLAS-CONF-2012-161} and
CMS \cite{CMS-PAS-HIG-13-012} can be summarized as follows. ATLAS reports a $95 \% $ C.L. limit on the signal strength
characterizing this process of $\{1.8, 3.4\}$, normalized to the SM,
 for operating energies of $\{7,8\} \, {\rm TeV}$ and assuming $m_h = 125 \, {\rm GeV}$. This compares to an expected limit of
$\{3.3, 2.5\}$ for these operating energies. Expressed in terms of extracted signal strengths $(\hat \mu)$ the reported values are
$\{-2.7 \pm 1.6,1 \pm 1.4 ,-0.4  \pm 1.1\}$ for $\{7,8, 7+ 8\} \, {\rm TeV}$ operating energies. Here we have added the reported error in quadrature. Similarly, CMS reports an observed $95 \% $ C.L. limit of $1.89$
compared to an expected limit of $0.95$ for $m_h = 125 \, {\rm GeV}$ using the combined $7+8  \, {\rm TeV}$ data set. The signal strength corresponding to the $2.1 \sigma$ local excess is reported to be $\hat \mu=1.0 \pm 0.5$ for $m_h = 125 \, {\rm GeV}$.

At present is not easy to translate these results into precise constraints on the form factors (or the EFT parameters).
In reporting a value for the signal strength in associated production, the
kinematics of the various events that are defined to correspond to the signal strength are averaged over. In particular, what is reported effectively averages over a value of $q^2$,
\bea
\hat{\mu} \equiv \int \, d q^2 \, R_V(q^2)  \times \mathcal{B}_b~,
\eea
for the events that pass the selection cuts. Here $\mathcal{B}_b = {BR_{BSM} (h \rightarrow \bar{b} \, b)}/{BR_{SM} (h \rightarrow \bar{b} \, b)}$. As can be seen by the from of  Eq.~(\ref{eq:Rcross}), the averaging over $q^2$ is particularly relevant for the sensitivity to different combinations
of the underlying parameters in the EFT. There is clearly more information in reporting $d \sigma/d \, q^2$ than in the averaged total signal strength. 

To see the sensitivity that the spectrum offers on the underlying parameters,  we normalize the average $\bar{q}^2$ of the 
$Vh$ final state as $\bar{q}^2 =  \mathcal{N} \, m_h^2 \, (1+ \sqrt{\rho})^2$.  Here $\rho = m_V^2/m_h^2$. There is an effective value of $\bar{q}^2$ that corresponds to a reported value of $\hat{\mu}$ for the total signal strength in associated production. Such a parameter is not reported by ATLAS and CMS, although clearly $\mathcal{N} \sim \mathcal{O} (1)$ and  $\mathcal{N} \geq 1$. We strongly encourage the experimental collaborations to report a value for $\bar{q}^2$ when any total associated production signal strength is reported in the future, and to bin the data in terms of different effective reconstructed $\bar{q}^2$ once their are sufficient events.

It is easy to illustrate the value of such a binning. Consider the case where the signal events are dominantly at threshold, so that 
$\mathcal{N} \approx 1$. This is the naive expectation, given the cross section falls rapidly due to the PDF 
 behavior. In this case, employing the form factor parameterization in Eq.~(\ref{eq:fNL}) and expanding to the 
 cross section to leading order in $g_2^2$,  we would obtain the following constraint
\be
 c_1^2 + 2 \, c_1 \, g_2^2 (c_2 + c_3) \left(1 + \frac{(1+ \sqrt{\rho})^2}{\rho} \right) + 4 \, c_1 \, c_3^\star \, g_2^2 \, \left(1+ \frac{1}{\sqrt{\rho}} \right)  = \hat \mu \, \mathcal{B}_b^{-1}~.
\ee
Conversely, considering a large $\mathcal{N}$ bin in associated production, would lead to 
\bea
 2 \, \mathcal{N} \, c_1 \, g_2^2 (c_2 + c_3) \frac{(1+ \sqrt{\rho})^2}{\rho}  = \hat \mu \, \mathcal{B}_b^{-1}~,
\eea
at leading order in $1/\mathcal{N}$.
As can be seen, a different set of Wilson coefficients is constrained in the cases $\mathcal{N} \sim 1$ and $\mathcal{N} \gg 1$.
Given the lack of information about the $\bar{q}^2$ corresponding to present data, and the fact the reported $\hat \mu$
values are still affected by $O(1)$ errors, in the following we will neglect the present experimental constraints.

\subsection{Future Associated Production Spectra}\label{spectra}

Eventually, with sufficient data, a binned associated production spectra as a function of $q^2$ can be experimentally constructed.
In this Section, we report some simple numerical examples of the importance of such spectra in constraining the underlying EFT.

\begin{figure}[t]
  \centering
  \includegraphics[width=6.5cm, height = 5cm]{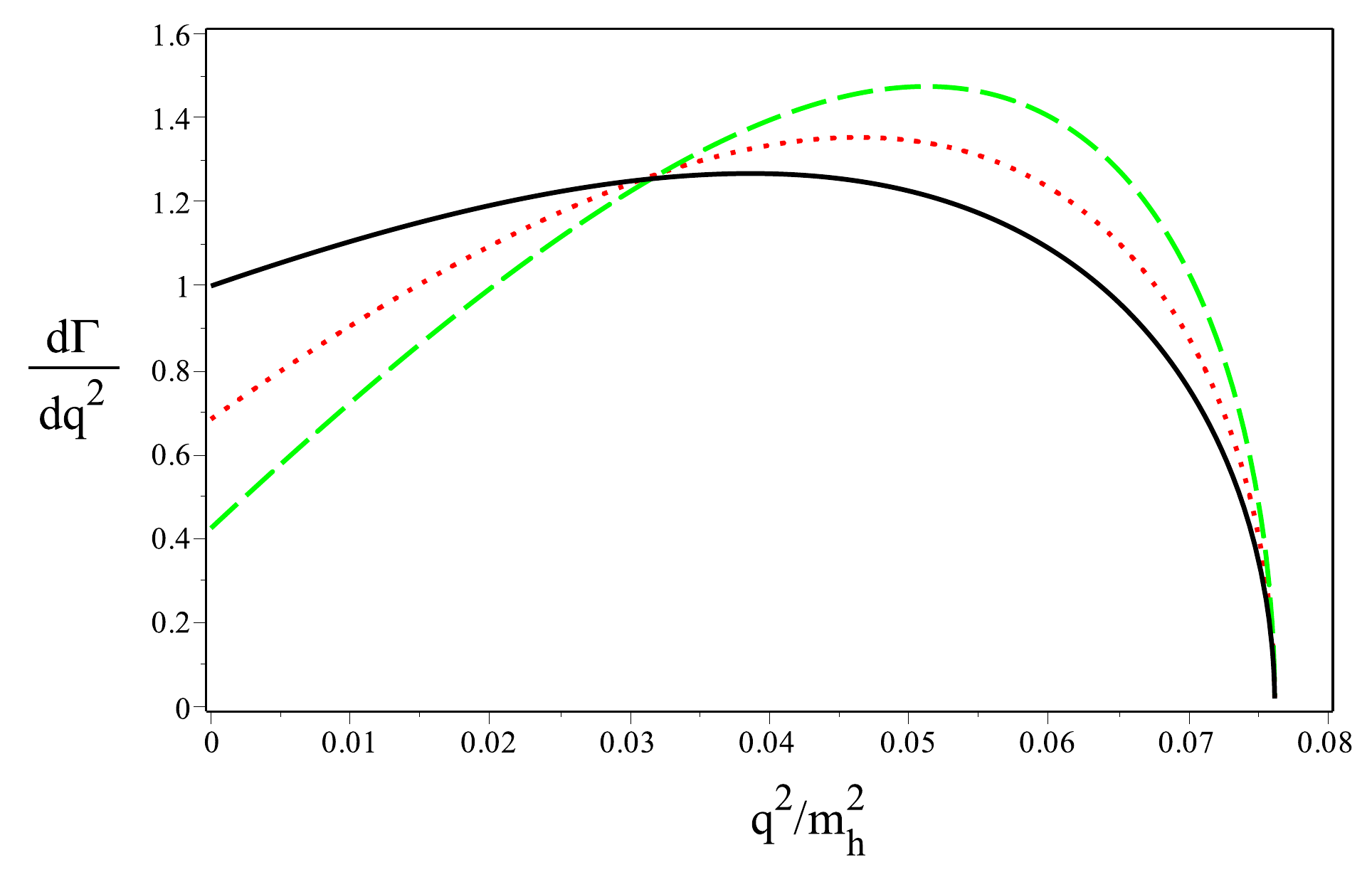}
 \includegraphics[width=6.5cm, height = 5.3cm]{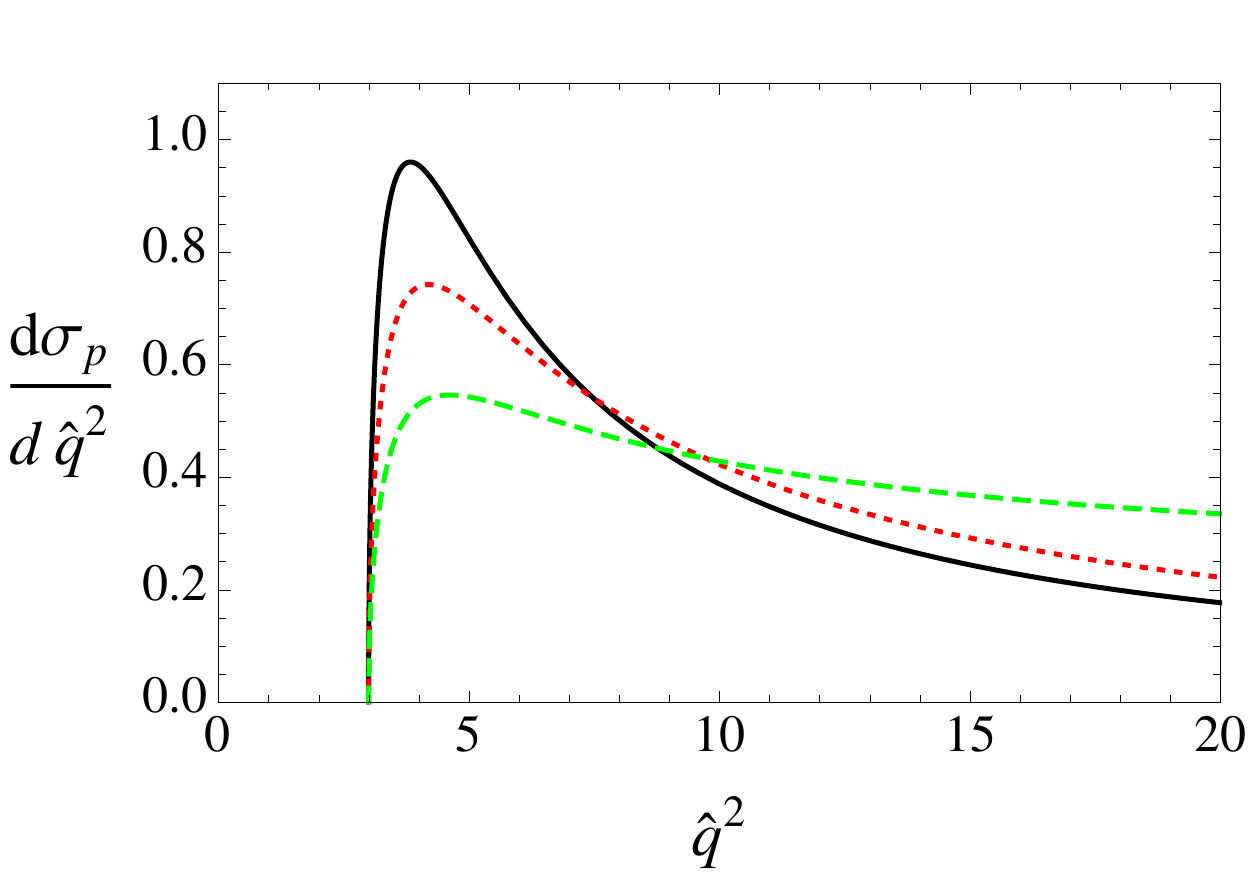}
  \caption{{\em Left:}  
  $\rd \Gamma(h\to Z\ell^+\ell^-)/\rd {\hat q}^2$ spectra ($\hat q^2=m_{\ell\ell}^2/m_h^2$),  in arbitrary units, for different values of the EFT parameters 
  chosen to leave the total  $h \rightarrow Z \ell^+ \ell^-$ rate unchanged. The values were not chosen so that the total integrated 
associated production cross section is the same. The black (full) curve corresponds to the SM, $c_i=(1,0,0)$, the red (dotted) curve is  for $c_i=(0.82,-0.8,0.8)$, 
  the green (dashed) curve for $c_i=(0.06,0,1.4)$. 
  {\em Right}: the partonic $d\sigma(q\bar q \to Zh)/d \hat{q}^2$  for the same EFT parameter choices (again arbitrary units). 
  For the sake of presentation, in the right curve the \{black, red, green\} curves  have been multiplied by a factor $\{ 5,1,2.7\}$.} \label{fig:spectra1}
\end{figure}

The kinematic distributions of associated production offer a complementary sensitivity to the unknown physics parameters. This is expected as the various
distributions probe different off-shell momentum regions, as previously noted.
In Fig.~\ref{fig:spectra1} (right) we show the effect of varying 
the form factors in the $Zh$ associated partonic production 
cross section at fixed $\hat q^2 = q^2/m_h^2$. We use the form factor parameterization in Eq.~(\ref{eq:fNL}) and we
restrict our detailed analysis to the case of the $Zh$ production, given the $q^2$ dependence of  $\sigma(\bar{\psi}_i \, \psi_j \rightarrow W h)$  
and  $\sigma(\bar{\psi}_i \, \psi_i  \rightarrow Z h)$  are qualitatively very similar.  In $Zh$ production, the start of the distribution,
corresponding at $\vec{p}_h = 0$,  occurs at the partonic value $\hat{q}^2  = (1 + \sqrt{m_Z^2/m_h^2})^2 \approx 3$.\footnote{~We use the numerical values $m_h = 125 \, {\rm GeV}$, $m_Z = 91.2 \, {\rm GeV}$, $\alpha_{\rm em}(m_Z)= 1/128.93$ and $\sin^2 \theta_W = 0.231$.}  For comparison, in Fig.~\ref{fig:spectra1} (left) 
the $\rd \Gamma(h \to Z \ell^+\ell^-)/\rd {q}^2$ spectra ($q^2=m_{\ell\ell}^2$) for the same set of  $c_i$ is also shown. 

The three sets of $c_i$ used in Fig.~\ref{fig:spectra1}  have been chosen such that they give rise to the same $\Gamma(h\to V\ell^+\ell^-)$ rate. 
In associated production the introduction of a non-SM $q^2$ dependence has a significantly larger impact, and it affects 
both the total cross section and the $d \sigma/ d {\hat q}^2$ spectrum.  
To better illustrate the difference in the $q^2$ distribution, in  Fig.~\ref{fig:spectra1}  we have adopted a different 
overall rescaling of the three associated production distributions, multiplying the  $\{black,red,green\}$ curves by  $\{ 5,1,2.7\}$. 
The corresponding total cross sections obtained after convoluting with PDF's\footnote{Normalizing our leading order result to the value of the total cross section as given by the Higgs xsection working group.} 
are $\sim \{0.4,2.3,0.9\} [\rm pb]$ for  $m_h = 125 {\rm GeV}$. The presence of a non-SM $q^2$ dependence is clearly evident 
already from the values of total cross sections, but the differential distribution provides independent information to decipher the underlying dynamics.

The extra sensitivity of the associated production spectra to possible non-SM $q^2$ dependence of the form factors 
comes at a cost in terms of predictivity of an EFT derivative expansion, if $c_{2,3}, |c_1-1|$ are of order one.
The general expression in Eq.~(\ref{eq:Rcross}) in terms of form factors makes no assumptions about the underlying EFT, 
and is valid for arbitrary values 
of $q^2$. However, the EFT expansion of the form factors in Eq.~(\ref{eq:fNL}) is based on a derivative expansion that
breaks down for $q^2 \sim v^2/|c_i-c_i^{\rm SM}|$, i.e. almost immediately above threshold if $c_{2,3},|c_1-1| \sim {O}(1)$.  This failure of the EFT expansion  is signaled by the increasing difference (at large $q^2$)
between the general decomposition in Eq.~(\ref{eq:Rcross}) --supplemented by the polynomial 
form factors in Eq.~(\ref{eq:fNL})--  and its truncated expression,
\bea
\label{scaling}
\frac{\sigma^{\rm EFT} (\bar{\psi} \, \psi \rightarrow Z h)}{\sigma^{\rm SM} (\bar{\psi} \, \psi \rightarrow Z h)} = c_1^2 + 2 \, g_2^2 \, c_1 (c_2 + c_3)\left(1 + \frac{q^2}{m_Z^2}\right) +  6 \, g_2^2 \, c_1 \, c_3  \left(\frac{q^2 + m_Z^2 - m_h^2}{\ph^2 + 3 \, m_Z^2}\right),\quad 
\label{eq:trunc}
\eea
where we have neglected all $q^4/m_V^4$ terms. 
For the illustrative $c_i$ adopted in Fig.~\ref{fig:spectra1} this difference exceed $50\%$ already 
for $q^2 > 5 m_h^2$ given we have chosen values of the $c_i$ with 
${O}(1)$ deviations from the SM. The large difference between truncated and non-truncated 
expressions for these sets of $c_i$ is shown by the comparison of upper-left and middle-left panels in Fig.~\ref{fig:spectra3}: 
we stress that none of the two panels can be considered as a realistic benchmark scenario, given higher-order terms are necessarily relevant.
This reinforces the need to perform the analysis in terms of
general form factors in case of sizable deviations from the SM already at low $q^2$ values. 
Conversely, the derivative expansion in terms of local operators
is predictive and under control for the more "natural" values of the Wilson coefficients 
adopted in the middle and right panels in Fig.~\ref{fig:spectra3},
as explicitly illustrated by the plots in the middle row of Fig.~\ref{fig:spectra3}.

It is also important to note that the partonic invariant mass distribution is not directly accessible in experiments at hadron colliders. Two key ingredients are necessary in order 
to provide a distribution  closer to what can be measured at the LHC: i) the convolution of the partonic cross section with the parton distribution functions (PDF's); ii) an estimate of how the error on the reconstructed final state four momenta limits the experimental sensitivity to the Wilson coefficients. We discuss each of these effects in turn in the following sections.

\subsection{Breit-Wigner smearing and experimental reconstruction uncertainty}

The experimental uncertainty on the reconstructed $p$ and $p_h$  four vectors 
have to be taken into account when accessing the experimental sensitivity of various distributions. We provide a first rough estimate of this 
uncertainty  by convoluting the theoretical distribution  with Breit-Wigner momentum distributions for $p$ and $p_h$, 
treating the corresponding $\Gamma_{h,V}$ not as a fundamental widths but as an experimental error associated with the momentum reconstruction. 
More explicitly, we first express the theoretical distribution as a function of  $p^2=m_V^2$ and $p_h^2=m_h^2$, 
we then convolute this distribution with appropriate Breit-Wigner for these two kinematical variables. 
In both cases we assume an effective experimental error leading to $\Gamma_{h,V} \sim {O}(10 \, {\rm GeV})$.

\begin{figure}[t]
  \centering
 \includegraphics[width=4.75cm, height = 5cm]{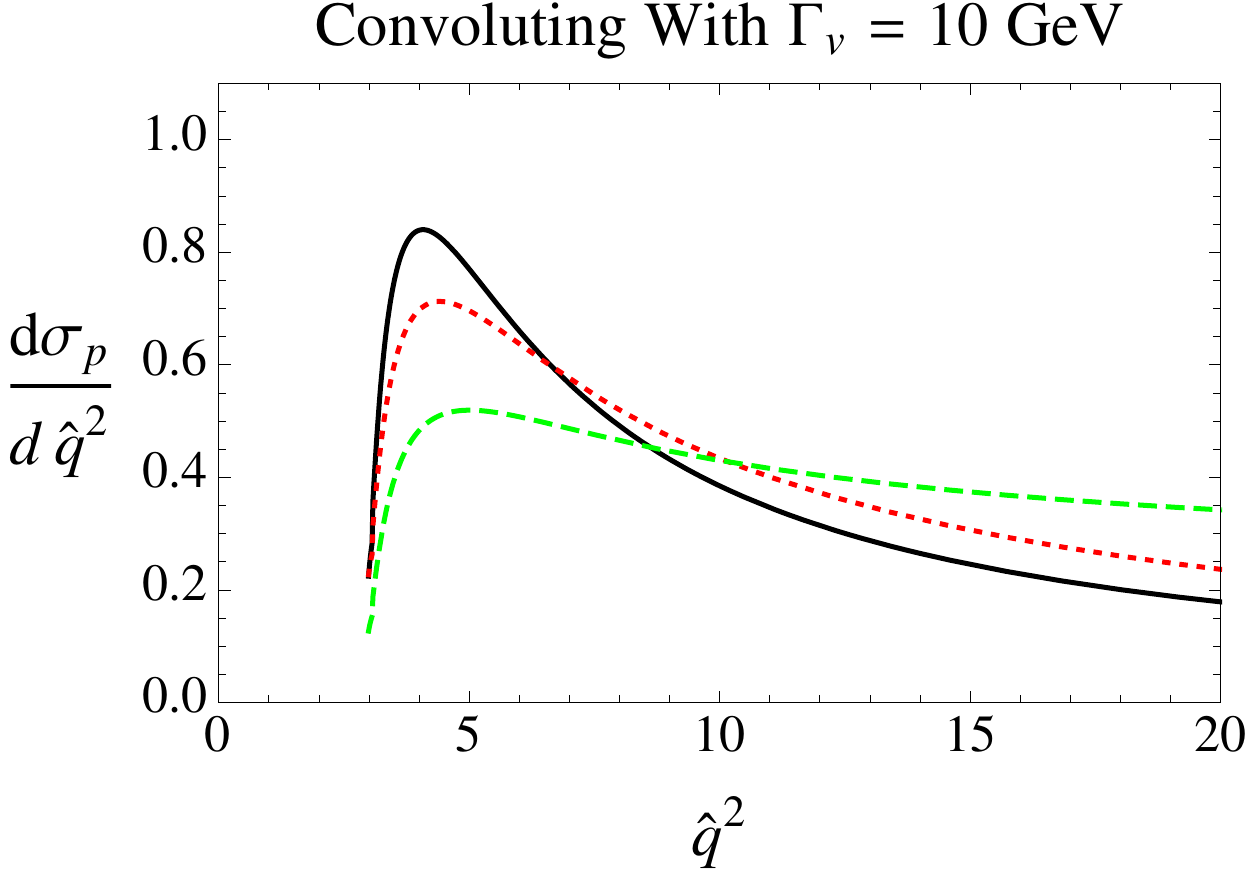}
  \includegraphics[width=4.75cm, height = 5cm]{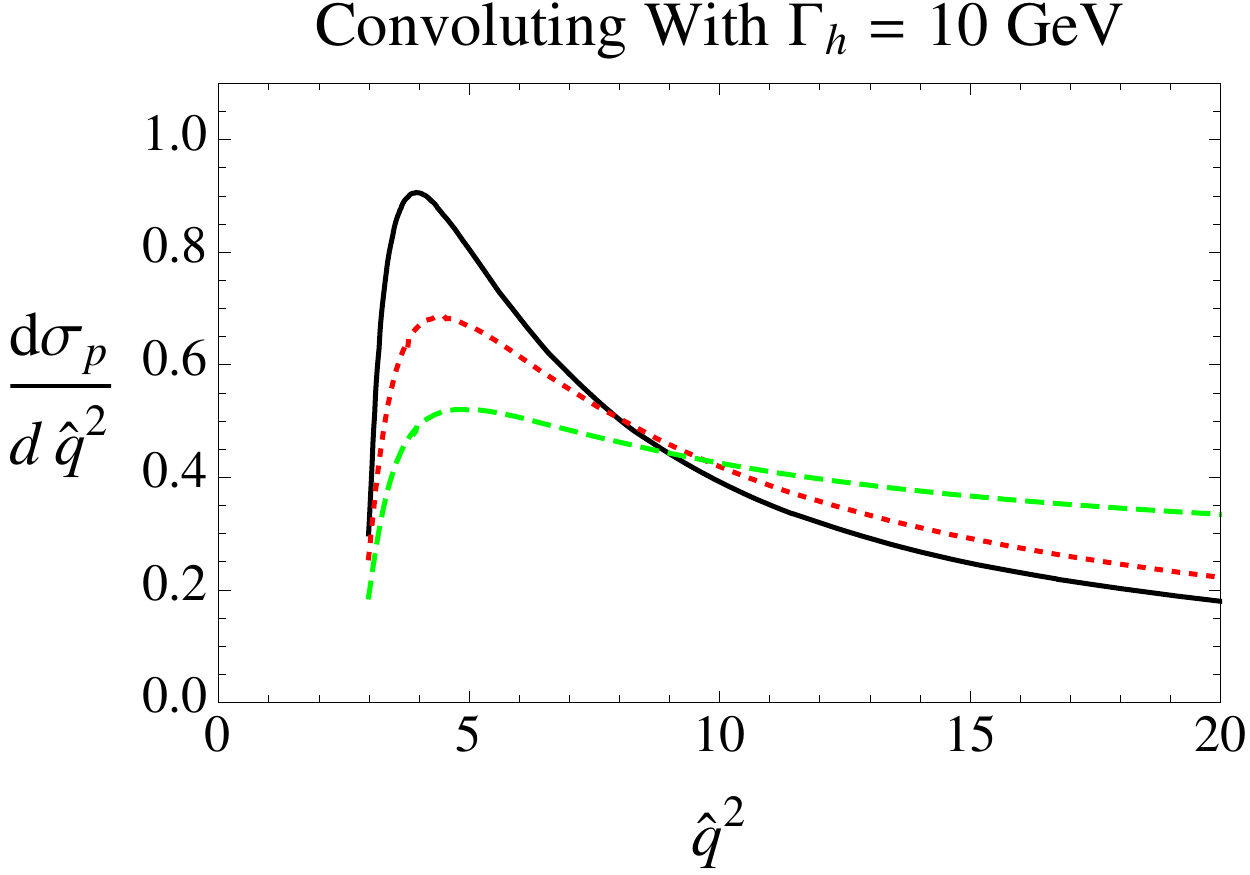}
  \includegraphics[width=4.75cm, height = 5cm]{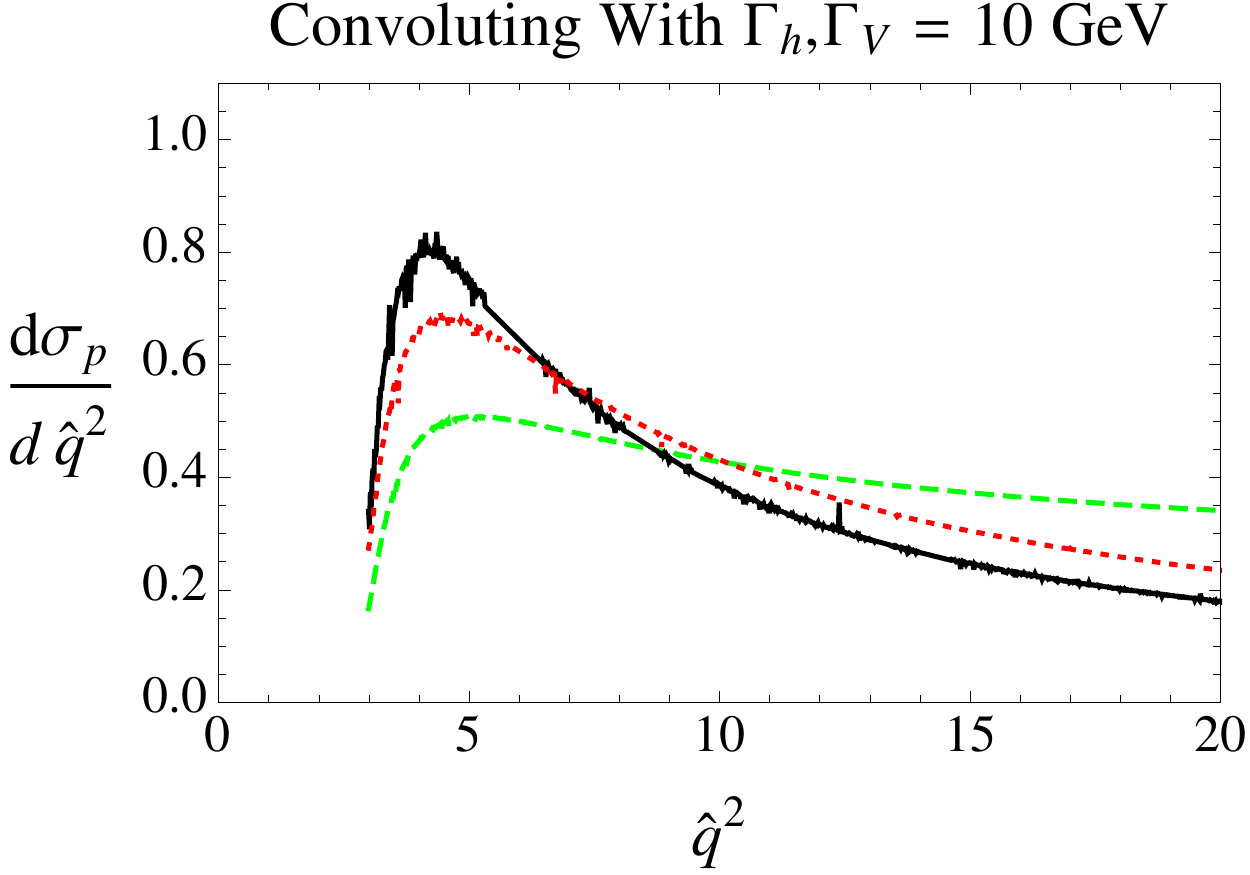}
    \caption{{\em Left}:  Convolution of the partonic $q^2$ spectrum with a Breit-Wigner distribution for the reconstructed on-shell $V$ with 
    an effective $\Gamma_V =10 \, {\rm GeV}$. 
  {\em Middle}: Convolution of the partonic $q^2$ spectrum with a Breit-Wigner distribution for the reconstructed $h$ with an effective $\Gamma_h = 10 \, {\rm GeV}$. {\em Right}: Convolution with a Breit-Wigner for both $V$ and $h$ invariant masses. 
  In all plots the color codes of the curves (and the corresponding normalization and parameter choice) is as in Fig.~\ref{fig:spectra1}.} \label{fig:spectra2}
\end{figure}

The effect of the experimental uncertainty is to suppress the peak in the threshold production region at low $q^2$.
Introducing the effective smearing discussed above, the effect is not particularly pronounced, as shown in Fig.~\ref{fig:spectra2}.
A more accurate modeling of the experimental resolution would be needed in order to provide a more quantitative 
assessment of this effect, but this preliminary investigation indicates that the sensitivity to the Wilson coefficients is not significantly degraded by these concerns. Note that errors in the reconstruction of $q^2$ itself are mitigated by a coarse binning of the reconstructed spectrum.

\subsection{PDF Effects, further distributions}

\begin{figure}[t]
  \centering
 \includegraphics[width=4.85cm, height = 5cm]{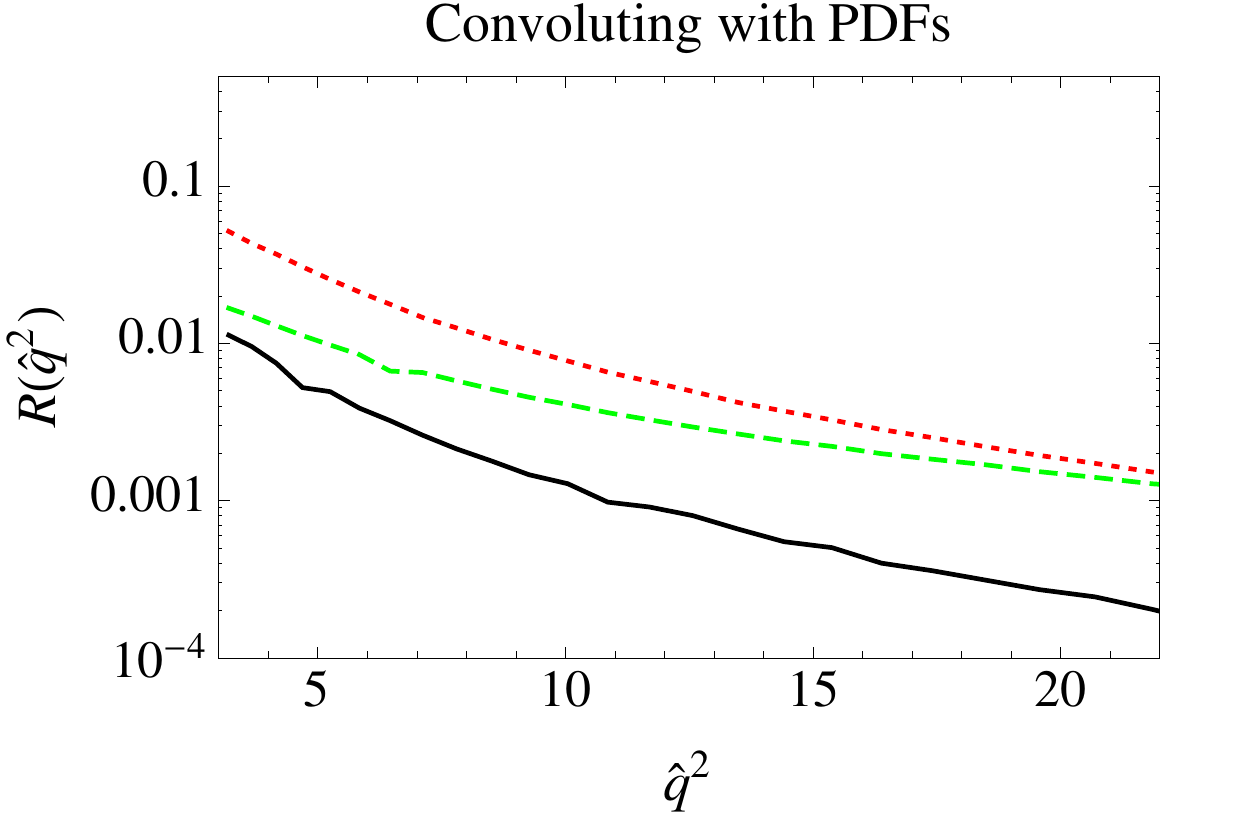}
  \includegraphics[width=4.85cm, height = 5cm]{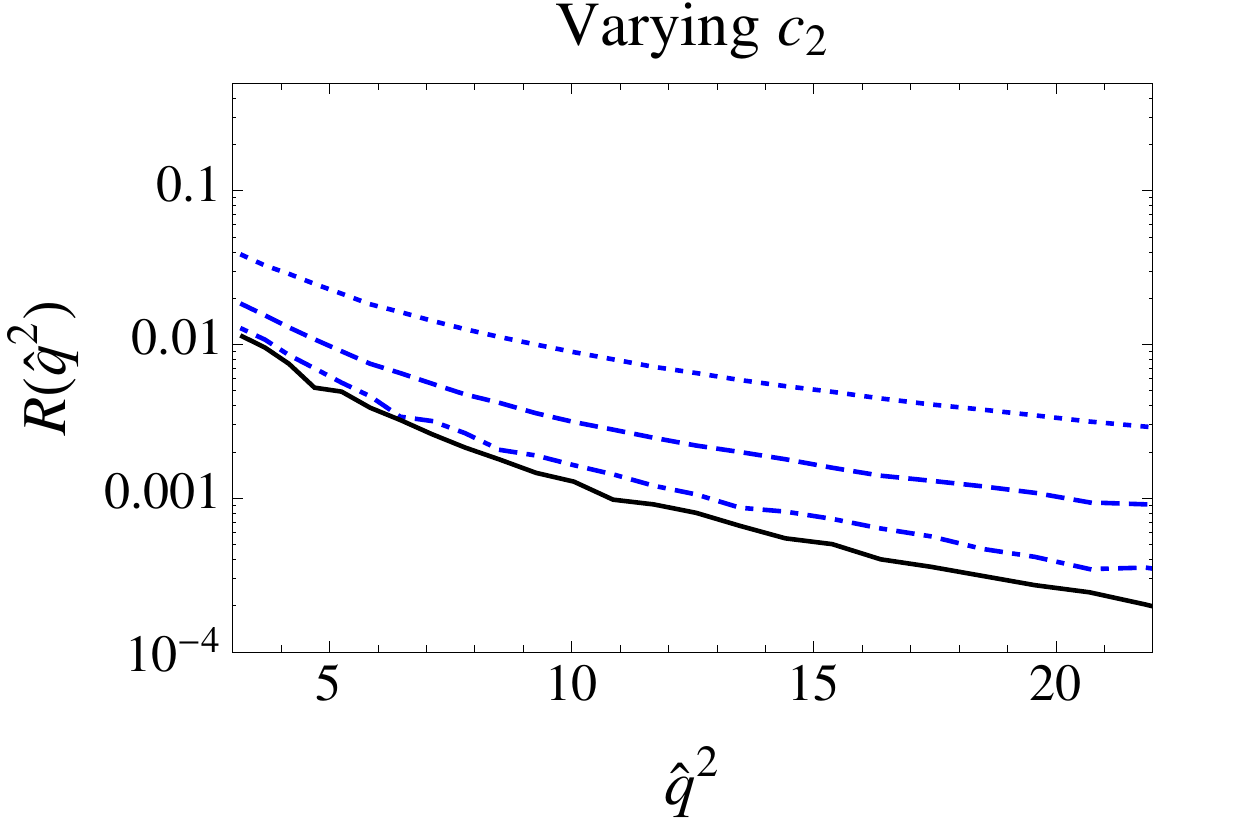}
 \includegraphics[width=4.85cm, height = 5cm]{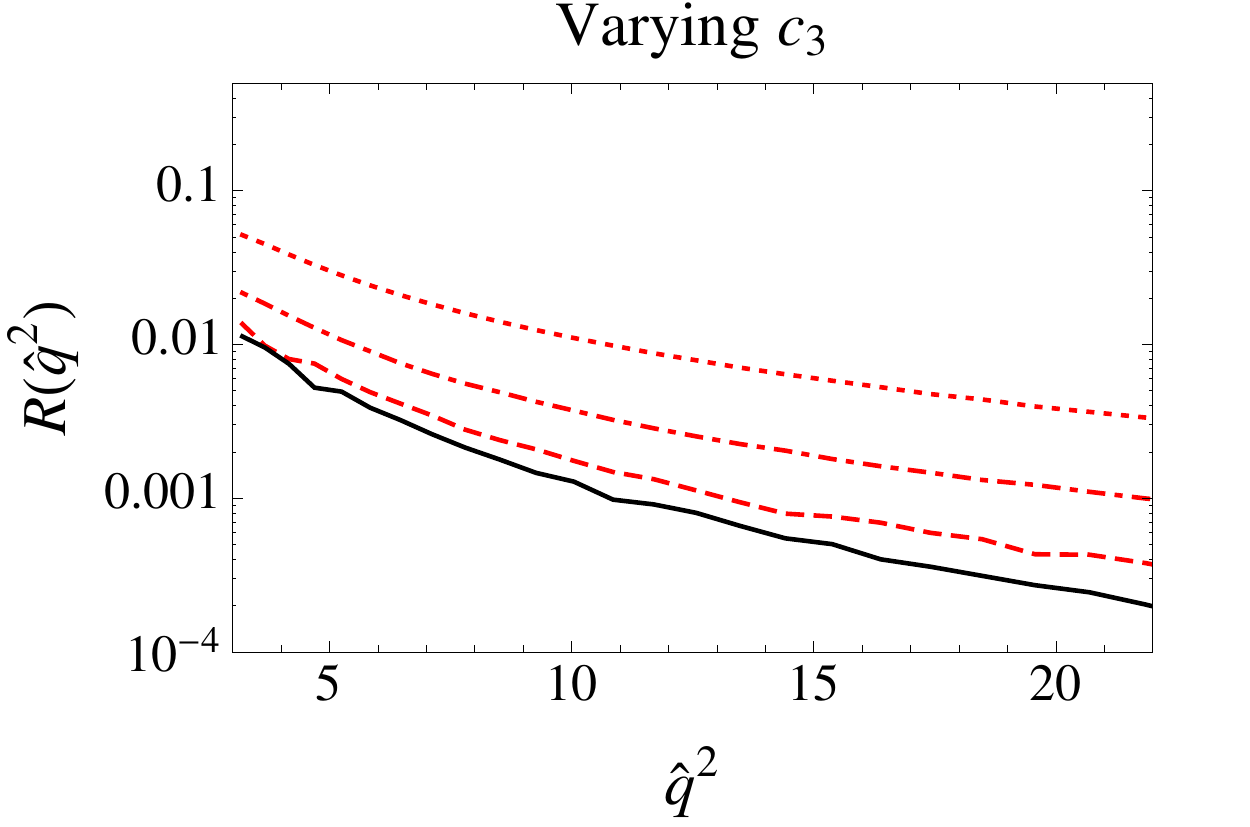}\\ 
 \vspace{0.25cm}
  \includegraphics[width=4.85cm, height = 5cm]{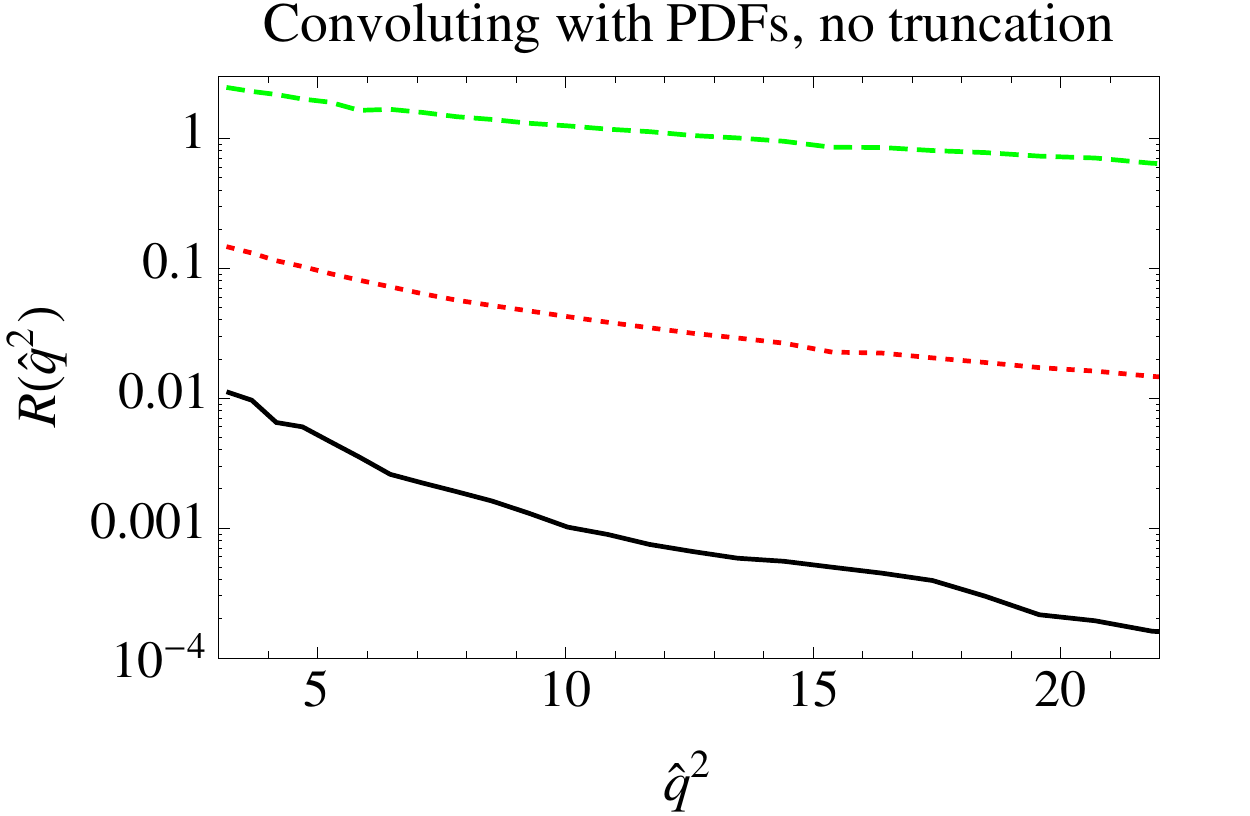}
  \includegraphics[width=4.85cm, height = 5cm]{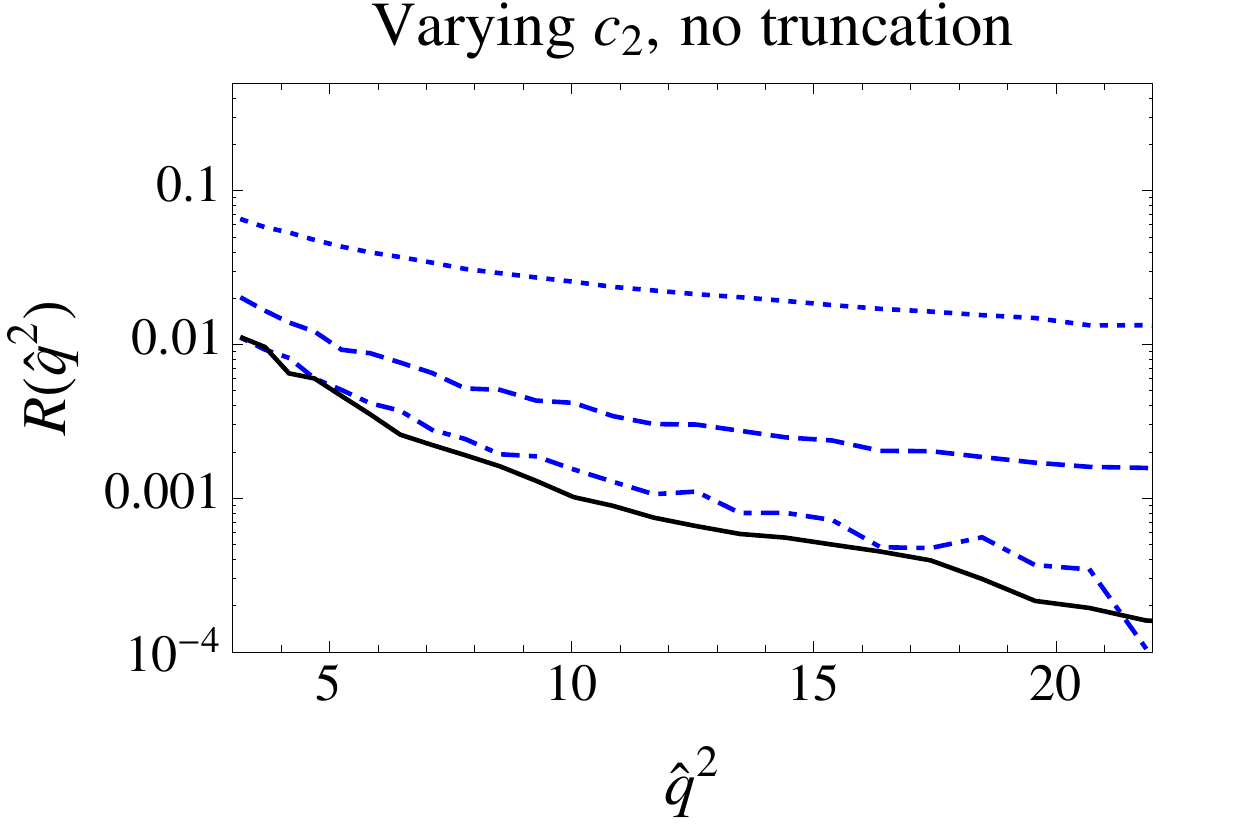}
 \includegraphics[width=4.85cm, height = 5cm]{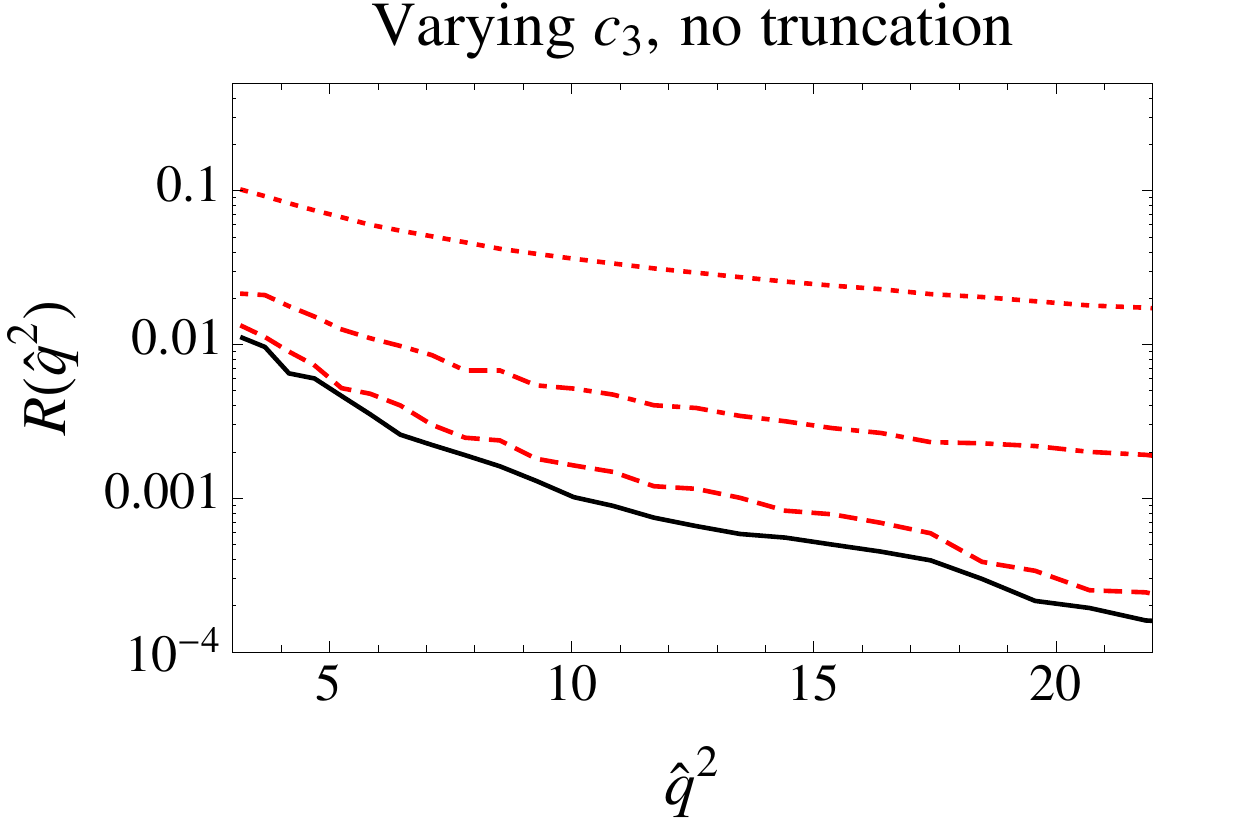}\\ 
 \vspace{0.25cm}
  \includegraphics[width=4.85cm, height = 5cm]{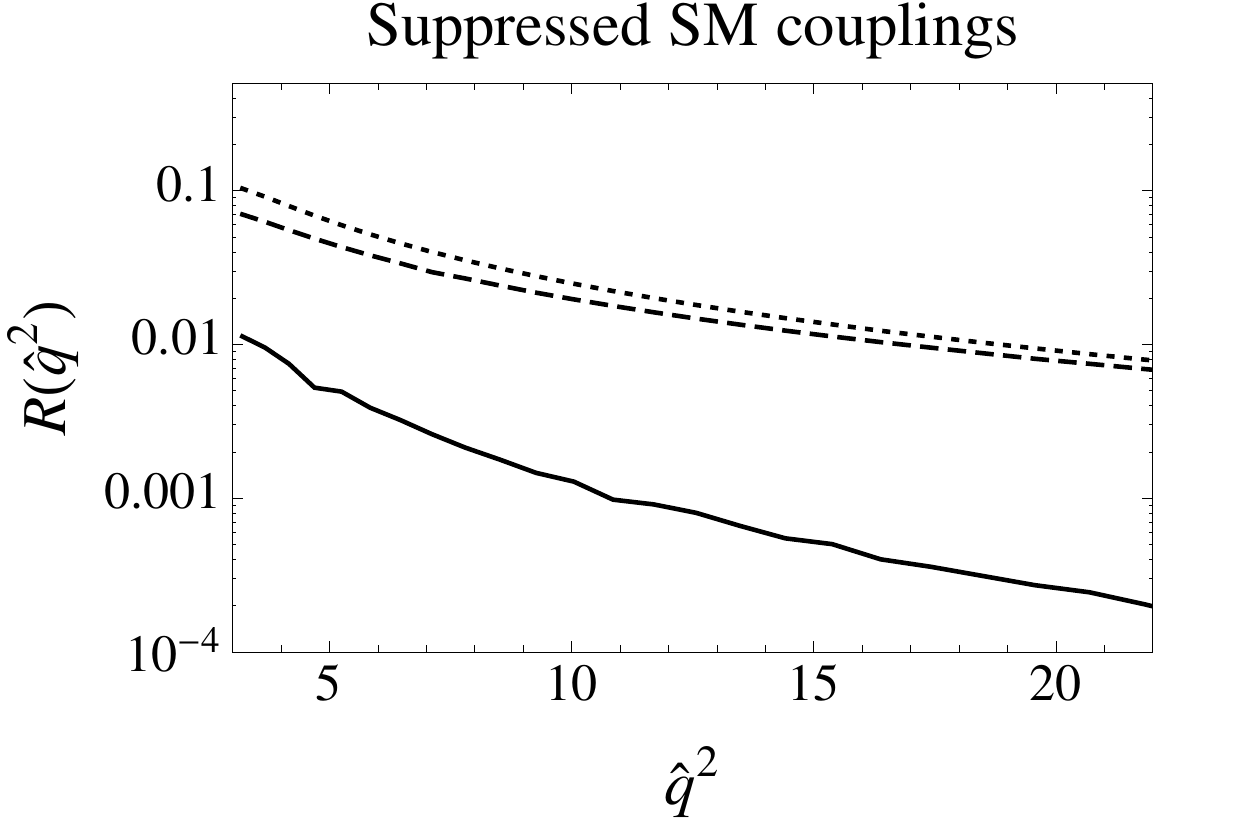}
  \includegraphics[width=4.85cm, height = 5cm]{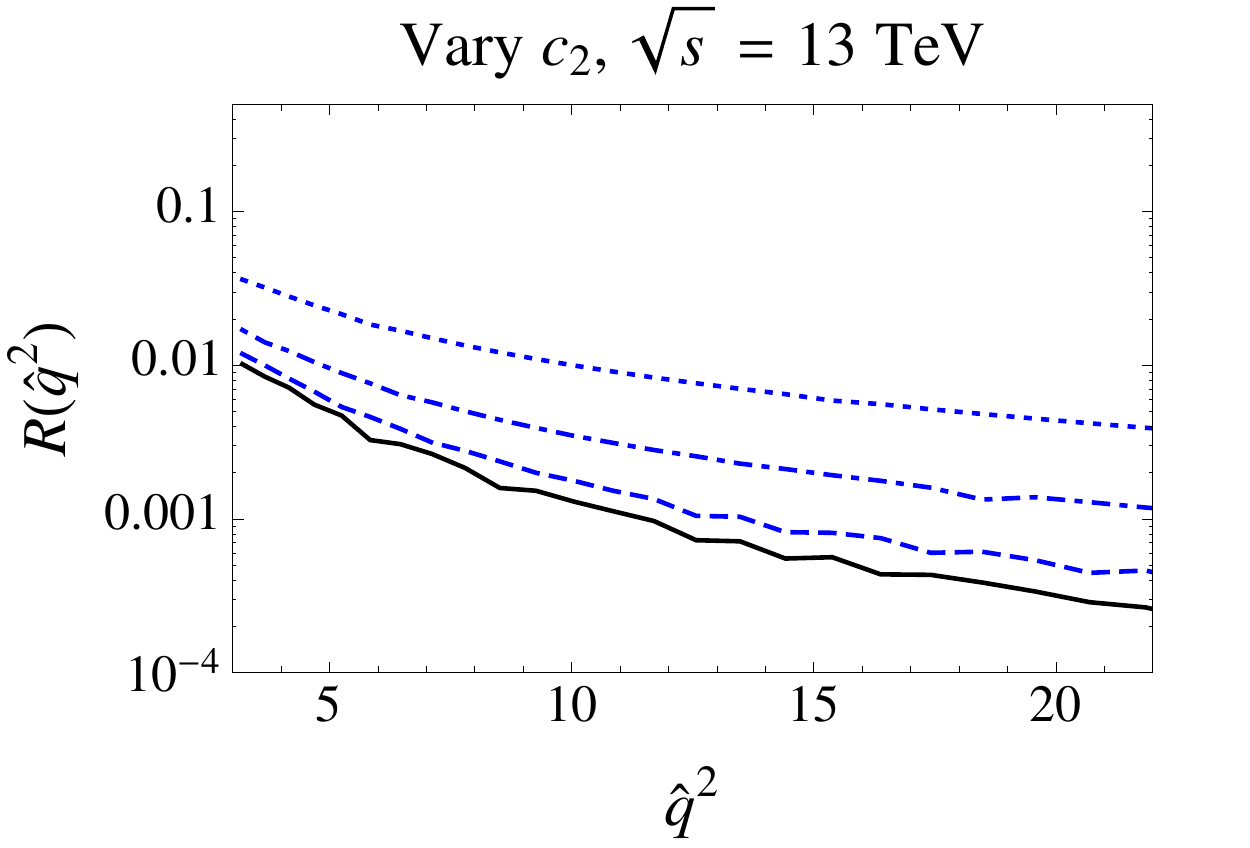}
  \includegraphics[width=4.85cm, height = 5cm]{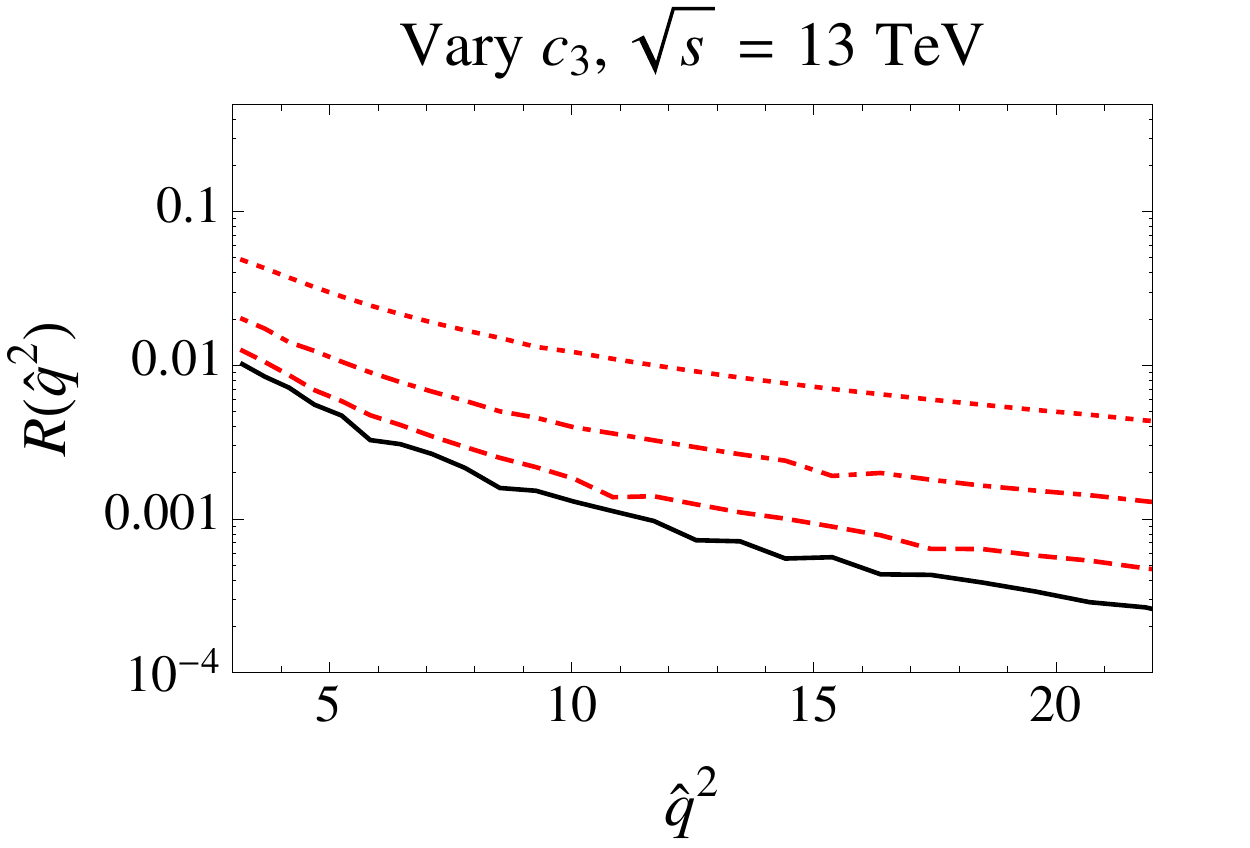}
  \caption{Differential $pp$ crossection normalized to the SM value: 
 $R(\hat q^2) =  [\sigma^{\rm SM}(pp\to Zh]^{-1} \times \times \rd \sigma^{\rm EFT}(pp \to Zh)/\rd {\hat q}^2$.
   {\em Top Left}:  $pp$ cross section for the same $c_i$ used in Fig.~\ref{fig:spectra1} ($\sqrt{s} = 8 \, {\rm TeV}$) using  Eq.~(\ref{eq:trunc})
. {\em Top Middle}: Varying the parameter $c_2$ over values $0.01$~(dashed), $0.05$ (dot-dashed) and $0.2$ (dotted) using Eq.~(\ref{eq:trunc}).  The parameters $c_1,c_3$ are fixed to $1,0$ in this case. {\em Top Right}: Varying the parameter $c_3$ over values $0.01$~(dashed), $0.05$ (dot-dashed) and $0.2$ (dotted) using Eq.~(\ref{eq:trunc}). The parameters $c_1,c_2$ are fixed to $1,0$ in this case.
 {\em Middle row}: Same as the top row except the un-truncated expression for Eq.~(\ref{eq:Rcross})  is used.
 {\em Bottom Left}: EFT parameters leading to a suppressed leading-order couplings of $h$ to the $Z$:
  $c_i=(0.5,  1,  0.01)$ for the dashed curve and $c_i=(0.5,  0.01, 1)$ for the dotted curve. {\em Bottom middle and right plot}: Same parameter choices as in the corresponding plots in the top row, for $\sqrt{s} = 13 \, {\rm TeV}$.} \label{fig:spectra3}
\end{figure}

We produce the hadronic invariant mass distributions for  $\sigma (pp \rightarrow Z h)$ 
by convoluting over MSTW2008 PDFs \cite{Martin:2009iq,Martin:2009bu}.
Due to the typical enhancement of the threshold region in the presence of anomalous couplings, the utility of reconstructing the
two body $d\sigma(q^2)/dq^2$ distribution is not significantly diminished by PDF effects. The effect of convoluting over the appropriate PDF's 
for the benchmark curves shown in Figs.~\ref{fig:spectra1}--\ref{fig:spectra2} is illustrated in Fig.~\ref{fig:spectra3}. 

Deviations in $c_1$ when
$c_2, c_3 \sim 0$ only normalize the expected SM distribution (that is re-scaled by $c_1^2$). In this case, since no extra momentum dependence is present, the differential distributions of associated production offer no advantage over precise measurements of the associated 
production signal strengths. However, in the presence of BSM expressed through
the effective couplings $c_{2,3}$, the non-SM momentum dependence in the form factor modifies the tail of the differential distributions substantially as shown in Fig.~\ref{fig:spectra3}. 
Further, the threshold region is typically significantly enhanced. Even low statistics reconstructions of these spectra would offer the opportunity to significantly constrain anomalous interactions
of a $0^+$ state in the general EFT. 

As shown in Fig. \ref{fig:spectra3},
the modification of the associated production cross section dependence on $q^2$ in the case of a nonzero  $c_{2,3}$ is highly degenerate. 
This is why disentangling the effects of anomalous $c_{2,3}$ couplings would most likely require a combination of studies of the effects in the associated production and in $h \rightarrow V \fstate$ decays. For illustrative purposes we also report the spectra corresponding to a 
suppressed leading-order couplings of $h$ to the $Z$, although this case is already ruled out by the  present 
ATLAS \cite{ATLAS-CONF-2012-161} and
CMS \cite{CMS-PAS-HIG-13-012} results. The 
plots shown here should be interpreted as illustrative of how large the BSM effects can be in 
the associated production spectra, providing a motivation for experimental studies aimed at reconstructing these spectra from data.

\section{Conclusions}\label{conclude}
In this paper we have examined the importance of using associated production spectra of the Higgs-like boson in constraining the properties of the newly discovered Boson $h$, that we have assumed to be  a $0^+$ state.
Combining the results of this paper with past work \cite{Isidori:2013cla}, we have systematically developed a formalism to decompose and characterize the
tree-point $h V \fstate$  Green's function, that can be probed both in $h$ decays ($h  \rightarrow V \fstate$) and  
in $Vh$ associated production ($pp \to \fstate \rightarrow h V$). 
Given the experimental value of $m_h$, what can be probed in such processes is only the on-shell $h V \fstate$  Green's function,
and not the $h V V$  one (that is kinematically forbidden).  This implies that, in generic extensions of the SM, it is necessary to 
 incorporate a possible non-standard momentum dependence in the form factor decomposition of  such processes. 
This non-SM $q^2$ dependence can make the differential spectra of these processes more sensitive probes of the
nature of the Higgs-like Boson than just the total signal strength.

Our results indicate that associated production is a powerful probe of the non-SM $q^2$ dependence that could be present in the $h V \fstate$ Greens function, offering enhanced sensitivity to non-SM properties of the newly discovered state. This is simply due to the fact that,
 by construction, in this process $q^2/m_V^2 \gg 1$
and the effect of a  non-SM derivative expansion for effective operators involving the $h$ field 
is enhanced.
We have also shown how, employing an EFT description of the form factors, 
low-$q^2$ and high-$q^2$ measurements of associated production constrain different combinations of  the 
underlying parameters in the EFT construction. We have demonstrated how to map the $h V \fstate$ form factors into 
EFT approaches based on a linear or a non-linear
realization of $\rm SU(2)_L \times U(1)_Y$. This has allowed us to clarify the importance of  $h  \rightarrow V \fstate$ and  
 $\fstate \rightarrow h V$ differential measurements as key probes to distinguish these two EFT approaches. 

When a total signal strength is reported for associated production, we emphasize to the experimental collaborations that it is essential to report a corresponding 
average $\bar{q}^2$ in order for the underlying EFT parameters to be properly constrained.
We also  strongly encourage the experimental collaborations to report a reconstructed $d \sigma/d q^2$ spectrum for differing values of $q^2$ in associated production as soon as sufficient data is collected to allow this reconstruction.
Even a low statistics reconstruction of this spectrum with a coarse binning is a key measurement to constraining the properties of the Higgs-like boson.

\appendix

\acknowledgments
We thank Aneesh Manohar for useful discussions and collaboration on related material. We also thank Gian Giudice and Witold Skiba
for useful discussions, and Jure Zupan for comments on the manuscript.
This research was  supported in part by the National Science Foundation under Grant No. PHY11-25915 at KITP Santa Barbara.
G.I.~acknowledges partial support by MIUR under project 2010YJ2NYW.

\bibliographystyle{JHEP}
\bibliography{workingnotes5}

\end{document}